\documentclass[aps,prx,reprint,amsmath,amssymb,superscriptaddress]{revtex4-2}

\usepackage{graphicx}
\usepackage{dcolumn}
\usepackage{bm}
\usepackage{comment}

\usepackage{hyperref}
\usepackage{color}
\usepackage{epsfig}
\usepackage{latexsym}
\usepackage{mathtools}
\usepackage{caption}
\usepackage{subcaption}
\usepackage{ulem}
\usepackage{verbatim}
\usepackage{url}
\begin{document}

\preprint{APS/123-QED}

\title{Topologically Protected Steady Cycles in an Ice-Like Mechanical Metamaterial}

\author{Carl Merrigan}
\email{carl.merrigan@fulbrightmail.org}
\affiliation{
 School of Mechanical Engineering, Tel Aviv University, Tel Aviv 69978, Israel
}
\author{Cristiano Nisoli}
\affiliation{
 Theoretical Division, Los Alamos National Laboratory, Los Alamos NM 87545, USA
}
\author{Yair Shokef}
\affiliation{
 School of Mechanical Engineering, Tel Aviv University, Tel Aviv 69978, Israel
}
\affiliation{
 Sackler Center for Computational Molecular and Materials Science, Tel Aviv University, Tel Aviv 69978, Israel
}

\date{\today}

\begin{abstract}
Competing ground states may lead to topologically constrained excitations such as domain walls or quasiparticles, which govern metastable states and their dynamics. Domain walls and more exotic topological excitations are well studied in magnetic systems such as artificial spin ice, in which nanoscale magnetic dipoles are placed on geometrically frustrated lattices, giving rise to highly degenerate ground states. We propose a mechanical spin-ice constructed from a lattice of floppy, bistable square unit cells. We compare the domain wall excitations that arise in this metamaterial to their magnetic counterparts, finding that new behaviors emerge in this overdamped mechanical  system. By tuning the ratios of the internal elements of the unit cell, we control the curvature and propagation speed of internal domain walls. We change the domain wall morphology from a binary, strictly spin-like regime, to a more continuous, elastic regime. In the elastic regime, we inject, manipulate, and expel domain walls via textured forcing at the boundaries. The system exhibits dynamical hysteresis, and we find a first-order dynamical transition as a function of the driving frequency. We demonstrate a forcing protocol that produces multiple, topologically-distinct steady cycles, which are protected by the differences in their internal domain wall arrangements. These distinct steady cycles rapidly proliferate as the complexity of the applied forcing texture is increased, thus suggesting that such mechanical systems could serve as useful model systems to study multistability, glassiness, and memory in materials. 
\end{abstract}

\maketitle

\section{Introduction}

Mechanical metamaterials are often created by compatible coupling of unit cells that  exhibit a floppy mode, so as to produce a desired global deformation~\cite{bertoldi2017flexible}. This design strategy has been employed to create materials exhibiting auxetic responses~\cite{bertoldi2010negative,shim2013harnessing,rafsanjani2016bistable}, collective pattern transformation~\cite{mullin2007pattern,kang2014complex,yang2016phase,rocklin2017transformable,ou2018kinetix,deng2020characterization}, multistable structures~\cite{coulais2018multi,jin2020guided}, violations of mechanical reciprocity~\cite{coulais2017static}, and simple programmability~\cite{florijn2014programmable,coulais2016combinatorial}. Any incompatibility in cell arrangement  implies geometric frustration~\cite{han2008geometric,shokef2011order,ortiz2016engineering}. Understanding how local frustration affects the global response is crucial, not only to avoid frustration which may hinder some functionality, but more interestingly for the controlled incorporation of frustration, which enables novel functionality~\cite{bertoldi2017flexible}. For example, taking periodic or nonperiodic compatible mechanical metamaterials~\cite{coulais2016combinatorial}, topological defects may be introduced by rotating individual unit cells, and are useful for steering stresses along designed trajectories~\cite{meeussen2020topological,meeussen2020response,pisanty2020topological}. Frustration between competing antiferromagnetic ground state orientations has been exploited to produce programmable hysteresis in perforated elastic sheets~\cite{florijn2014programmable}, and frustration was found to generate multiple complex chiral patterns in triangular networks of beams~\cite{kang2014complex}. In general, metamaterials incorporating frustration can often accommodate the conflicts in surprising ways, inducing complex and potentially useful mechanical responses.

The design of new properties emerging from frustration~\cite{nisoli2018frustration,nisoli2017deliberate} has also been extensively studied in engineered magnetic nanomaterials based on the ice rule~\cite{bernal1933theory,Pauling1935}, and called artificial spin ice (ASI)~\cite{nisoli2013colloquium,skjaervo2019advances,ortiz2019colloquium}. ASI can be helpfully modeled as sets of classical Ising spins arranged along the edges of a lattice, where each spin lies in the plane of the lattice and points away from one vertex and towards another~\cite{baxter1980exactly}. It is possible to map an ASI to a mechanical metamaterial in the following way (see Fig.~1): We map each vertex of the ASI to a mechanical unit cell which has a single floppy mode. Importantly, the mechanical floppy mode is such that the deformation of each unit-cell edge may be mapped to a binary spin within a ground-state ASI vertex. Neighboring cells that attempt to adopt incompatible floppy modes create local stress, raising the energy of the metamaterial. Thus, the rules for reducing or manipulating frustration in ice-like systems relate to the rules of stress compatibility in metamaterials. However, we emphasize that the latter are still continuum, elastic systems. Therefore, we expect that mechanical spin-ice systems, unlike their magnetic counterparts, will accommodate the effects of frustration in ways that are not afforded to a truly binary system, leading to novel phenomenology.

In this paper, we put forward the first step in a broader program to relate concepts and models between spin systems and mechanical metamaterials. Here we begin with the simplest geometry---a two-dimensional mechanical analogue of square ice~\cite{morgan2011thermal,budrikis2012domain,farhan2013direct}. First we explore static configurations of the metamaterial, which we show may be tuned from a binary, pseudo-spin behavior to a continuous response. Then we go on to analyze the response to external driving. A novel and crucial property of our mechanical metamaterial is the possibility of manipulating it by acting on its boundary. Magnetic systems are typically driven by using a bulk magnetic field~\cite{tome1990dynamic,rao1990magnetic,chakrabarti1999dynamic,berglund1999memory,korniss2000dynamic} which acts on all the spins at once. In contrast, for our mechanical system, it is natural to push and pull on the faces of the unit cells at the metamaterial boundaries. When driven in a compatible way at the boundaries, our mechanical metamaterial exhibits a {\it dynamical phase transition} from asymmetric to symmetric hysteresis loops, as a function of the driving frequency, as well as the existence of a wide variety of distinct, topologically-protected steady cycles under textured driving protocols. This nontrivial dynamics arises from the topology and motion of domain walls in the system and could be developed as a dynamical form of memory~\cite{keim2019memory,Ido2019,Ido2020}.      
\section{Mechanical Spin Ice}

\begin{figure}[t!]
\begin{subfigure}[c]{\linewidth}
\centering
\resizebox{0.9\linewidth}{!}{\includegraphics{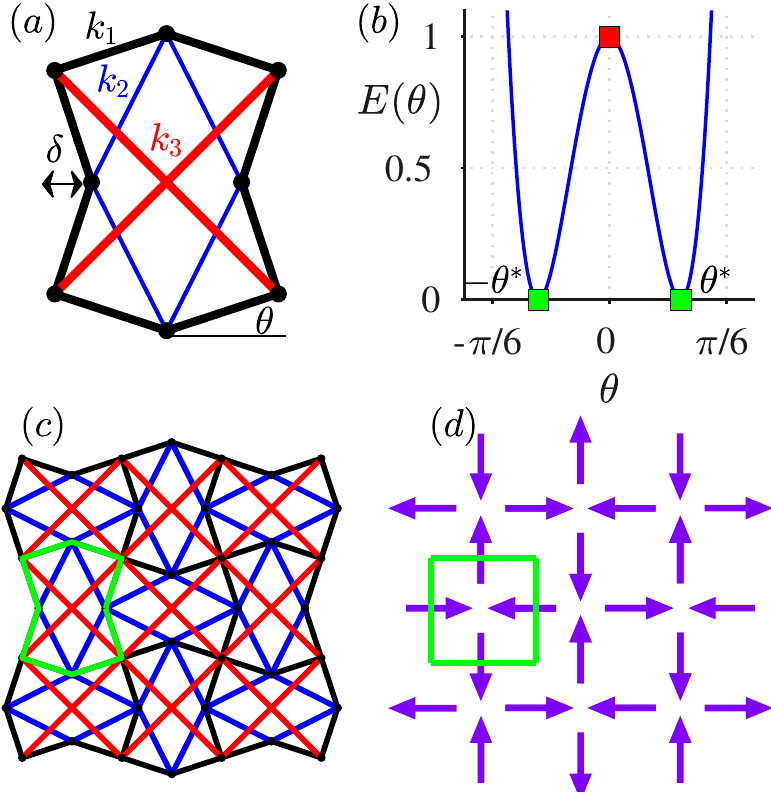}}
\end{subfigure}
\hfill
\caption{Bistable mechanical unit cell and ground state configurations of the elastic network. (a) Square mechanical cell composed of four right triangles. The cell is constructed from eight framework ($k_1$,black), four interaction ($k_2$,blue), and four compressive ($k_3$,red) springs with stiffness $k_1$, $k_2$, and $k_3$ and relaxed lengths $l_1$, $l_2=\sqrt{2}l_1$, and $l_3<2\sqrt{2}l_1$, respectively. (b) Energy as a function of the right triangle rotation angle $\theta$ for $l_3/l_1 = 0.95 \times 2\sqrt{2}$. The undeformed state (red) and the two ground states (green) are denoted by solid squares. (c) Ground state of a $3\times3$ network. A single unit cell is highlighted in green. (d) Mapping of the unit cell displacements to binary spins within square ASI.}
\end{figure}

We design elastic networks to mechanically mimic square ice. Four rigid right triangles, combined as a square, exhibit a floppy mode in which all the triangles may freely rotate together by some arbitrary angle $\theta$ without stretching any bonds [see Fig.~1(a)]. This general rigid mechanism is the basis for a variety of metamaterial designs~\cite{bertoldi2017flexible}, such as auxetic and chiral metamaterials~\cite{shim2013harnessing,rafsanjani2016bistable,ou2018kinetix}, metamaterials with propagating internal domain walls~\cite{deng2020characterization}, and also pattern-transforming perforated elastic sheets~\cite{mullin2007pattern,florijn2014programmable,yang2016phase}. If the face nodes where the triangles meet are treated as pseudo-spins, then this global mechanism maps to the ground states in magnetic ASI [Fig.~1(d)], for which each vertex has two opposing spins pointing inwards and two spins pointing outwards. This ``ice-rule''  constrains the ground states of spin ices to compatible combinations of vertices with zero topological charge, that is, the number of spins pointing towards the vertex equals the number of spins pointing away from the vertex~\cite{nisoli2013colloquium}.

Our square cell is composed of three types of springs [Fig.~1(a)]. Two are standard in assuring the floppy modes: ``Framework springs'' of stiffness $k_1$ and relaxed length $l_1$ form the boundary of the square, while ``interaction springs'' with stiffness $k_2$ and relaxed length $l_2 = \sqrt{2}l_1$ connect the four faces, completing the four right triangles. In ASI, spins are strictly binary, with magnetization per spin that can point in two possible directions but which is fixed in magnitude~\cite{nisoli2013colloquium,skjaervo2019advances}. However, in mechanical metamaterials the pseudo-spin magnitudes may vary continuously, and in typical designs the undeformed state ($\theta = 0$ in Fig.~1(a)) with all spins equal to zero is the unique ground state. Furthermore, deformations may continuously vary over large length scales, so that crisp, topological-defect structures do not appear spontaneously. Consequently, usually frustration comes into play only when a mechanical system is driven~\cite{coulais2016combinatorial,meeussen2020topological,meeussen2020response,deng2020characterization}. To obtain bistability, we introduce diagonal ``compressive springs'' with stiffness $k_3$ and length $l_3$. By requiring $l_3/l_1 < 2\sqrt{2}$, the square cells experience a local compressive stress that causes them to act as bistable {\it hysterons}~\cite{keim2019memory}, with two states $\pm\theta^*$ (Fig.~1(b)). Each edge of the square cell prefers to point in or out by some fixed magnitude, causing the overall metamaterial to mimic a spin ice with discrete states. In order for the compressive springs and the framework springs to simultaneously achieve their relaxed lengths, the geometry determines the deflection angle to be $\theta^* = \cos^{-1}{[l_3/(2\sqrt{2}l_1)]}$. 
 
{\color{black} We consider square mechanical metamaterials composed of $L_x$ columns and $L_y$ rows of bistable square cells, and we take $L_x=L_y$. The} low-energy, metastable configurations of this metamaterial are governed by tuning the stiffnesses of the three spring types used to construct the unit cell, of which one can choose two independent dimensionless ratios. For simplicity, we adopt units such that $k_1 = 1$, so that the values of $k_2$ and $k_3$ indicate the dimensionless relative stiffness of the interaction and compressive springs compared to the framework springs.

\begin{figure*}[t!]
  \includegraphics[width=1\textwidth]{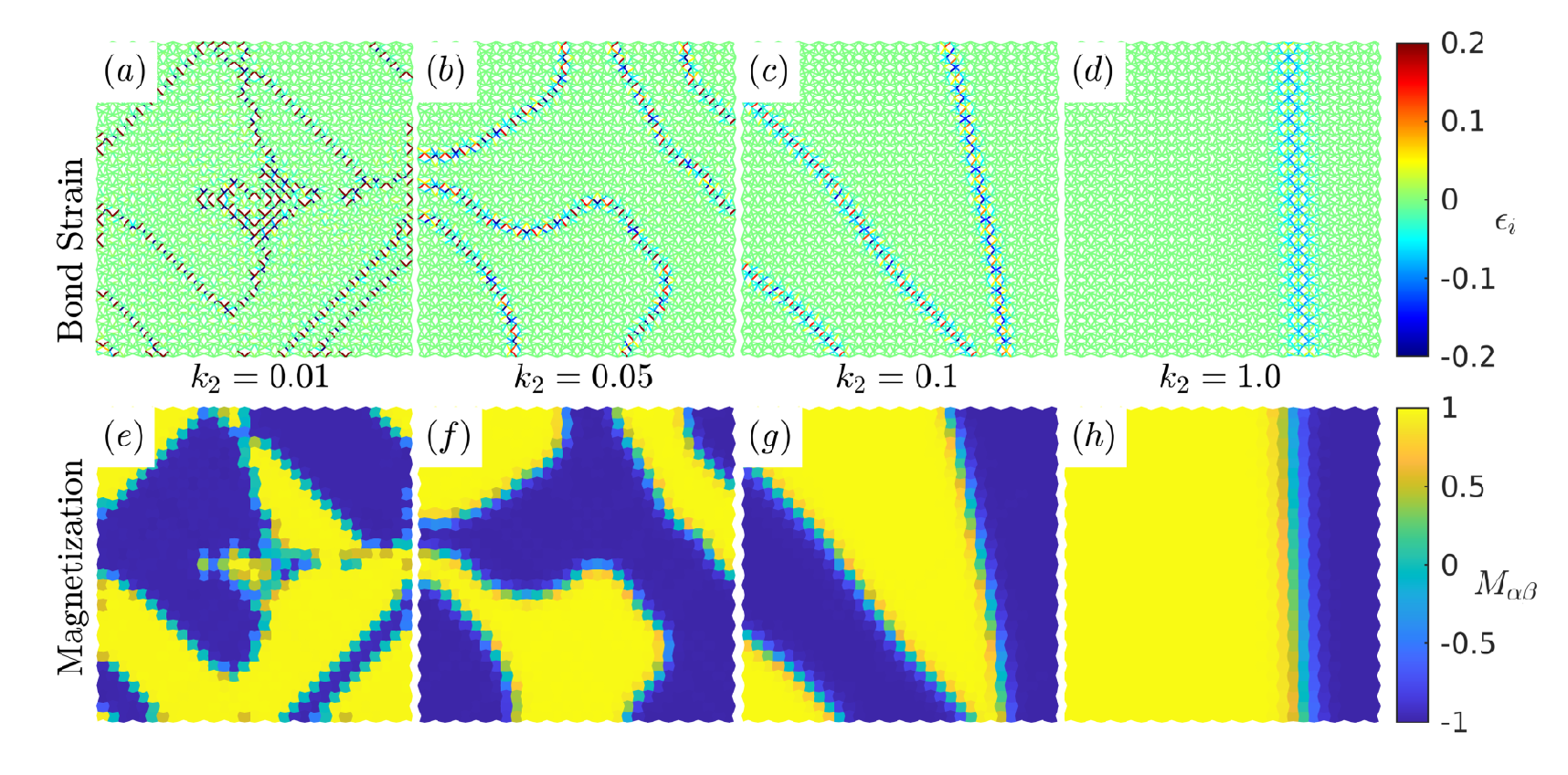} 
\caption{Defect and domain wall morphology of metastable states quenched with free boundary conditions as a function of the dimensionless stiffness $k_2$ of the interaction springs for lattices size $L_x = L_y = 30$. The compressive dimensionless stiffness is set to a large value $k_3=5$ for all cases. (a)-(d) Strains in the network for the framework and interaction springs (not shown are the diagonal compressive springs which are all nearly relaxed). (e)-(h) Staggered magnetization $M_{\alpha \beta}$ for the corresponding configurations. As $k_2$ is increased, the domain wall curvature is reduced and the gradients in the magnetization along the boundary between competing orientations become more smooth.}
\end{figure*}
	
As in the ice-rule obeying  ASI, we have two ``antiferromagnetically'' ordered ground states [Fig.~1(c)] in which  cells alternate orientations. We quantify the order in the system using the following {\it staggered magnetization}: 
\begin{equation} 
M_{\alpha\beta} = (-1)^{\alpha + \beta}\frac{(\Delta x_{\alpha \beta} - \Delta y_{\alpha \beta})}{4 \delta}, \, \alpha = 1...L_x, \beta = 1... L_y,
 \end{equation} 
 where $\Delta x_{\alpha \beta} $ is the width of the unit cell, $ \Delta y_{\alpha \beta} $ is its height, and $\delta = l_1sin(\theta^*) $ is the displacement amplitude of the face nodes in the ground state [Fig.~1(a)]. $M$ has value $1$ on all the cells of one ground state, $-1$ on the other, and $0$ on undeformed cells, and might be greater than $1$ under elastic deformation. Then, the globally averaged magnetization is  
 \begin{equation} 
 M(t) = \frac{1}{L_x L_y} \sum_{\alpha = 1}^{Lx} \sum_{\beta = 1}^{L_y} M_{\alpha \beta}(t). 
 \end{equation} 

 \section{Metastable Configurations}
 
We numerically simulate our metamaterial within an overdamped dynamics, subsuming any physical viscosity in our definition of time, see Appendix A. In the numerical simulations shown below we use $l_3/l_1 = 0.95 \times 2 \sqrt{2} $, so that $\theta^* = 18.2^{\circ}$ and $\delta/l_1 = 0.31 $.

Starting from a neutral state $M = 0$ where all cells are square, the system relaxes into metastable configurations corresponding to domains of different orientation of $M_{\alpha \beta}$, separated by domain walls, see Fig.~2. Different low-energy, metastable configurations of the metamaterial are obtained when uniformly distributed, random perturbations  $ \Delta x_i, \Delta y_i \in [-\chi,\chi]$ are applied  to each node of the lattice (here $\chi/l_1 = 0.4$). Distinct minima are then found by repeating the simulation for different random initial conditions. 

By gauging the dimensionless interaction stiffness $k_2$, we can control how the metamaterial mimics its binary counterpart and therefore whether its excitations  match the topologically charged vertex configurations characteristic of ASI~\cite{morgan2011thermal,budrikis2012domain,farhan2013direct}. Moreover, with our continuous, elastic system, we can also investigate regimes that go beyond strictly binary behavior, evincing new phenomena that would be absent in magnetic ASI.

First, consider $k_3 \gg  1 \gg k_2 $. The large value of the compressive stiffness $k_3$ constrains the corners of each unit cell to move inwards, so that the sides of the square unit cell have length less than $2l_1$. Therefore, in order to relax the framework springs, the four nodes at each face need to buckle in or out with amplitude $ \pm \delta$. For such configurations, only the interaction springs are strained: applying a spin approximation, $2^4$ configurations of the unit cell are possible (see Appendix B,~Fig.~7). The ice-rule is obeyed for two of these configurations and they correspond to the energy minimum, as in Fig.~1(b). The ice-rule is violated by excitations for which two neighboring faces both point inwards or outwards, causing the interaction springs to be strained. We note that with large $k_3$ compressing the interaction springs costs more energy than extending them, and this alters the vertex energy hierarchy compared to ASI (see Appendix B). When, $k_2 \sim 1 $, the pseudo-spin approximation breaks down and a wider variety of continuous excitations may appear in which stresses are shared between both the framework springs and the interaction springs. 

The dependence of the excitation morphology on the dimensionless interaction stiffness $k_2$ is demonstrated in Fig.~2. The compressive stiffness is large, with value $k_3 = 5$ throughout, and $k_2$ varies from $0.01$ to $2$. An experimental realization would require component springs with a variable stiffness range of order $500:1$, which may readily be achieved with standard materials. For a small dimensionless stiffness $k_2 = 0.01$ of the interaction springs, the domain walls have shapes very similar to ASI walls: one cell thick, and made of configurations in which two neighboring faces of the same cell point in the same direction~\cite{cristiano2020topological}. We also see other topological defects, especially at the edges of the lattice, corresponding to configurations in which three faces deform in the same direction, and are analogous to ASI monopoles (see Appendix B,~Fig.~8). This suggests that such topological charges may be expelled toward the boundaries of the system, similarly to what was shown for colloidal ice~\cite{libal2006realizing,ortiz2016engineering,libal2018ice,ouguz2020topology}. As $k_2$ increases, such ``monopoles'' become more energetically costly and appear within the domain walls. The curvature of the latter grows smaller with increasing $k_2$. Domain walls seen in Fig.~2(b), with $k_2 \sim 0.05$, resemble domain walls seen in experiments on square, magnetic ASI~\cite{morgan2011thermal,budrikis2012domain}. As $k_2$  increases further, the domain walls become straight, and they become wider with strains distributed over neighboring rows or columns of unit cells.

During the energy quench, all domain walls tend to move and straighten for some time until stabilizing at some average curvature imposed by the value of $k_2$. Further, any domain walls that connect two adjacent edges of the lattice and start near a corner tend to be expelled. Corner spanning domain walls are expelled at a decreasing speed as $k_2$ is decreased, until $k_2 \sim 0.18$, where the corner spanning domain walls become stationary and are no longer expelled (see Appendix C, Fig.~9). At values of $k_2$ which approach $1$, corner spanning domain walls tend to be eliminated rapidly, and the resulting configurations only contain straight domain walls, such as in Fig.~2(d). 

\begin{figure}[t!]
\begin{subfigure}[c]{\linewidth}
\centering
\resizebox{\linewidth}{!}{\includegraphics{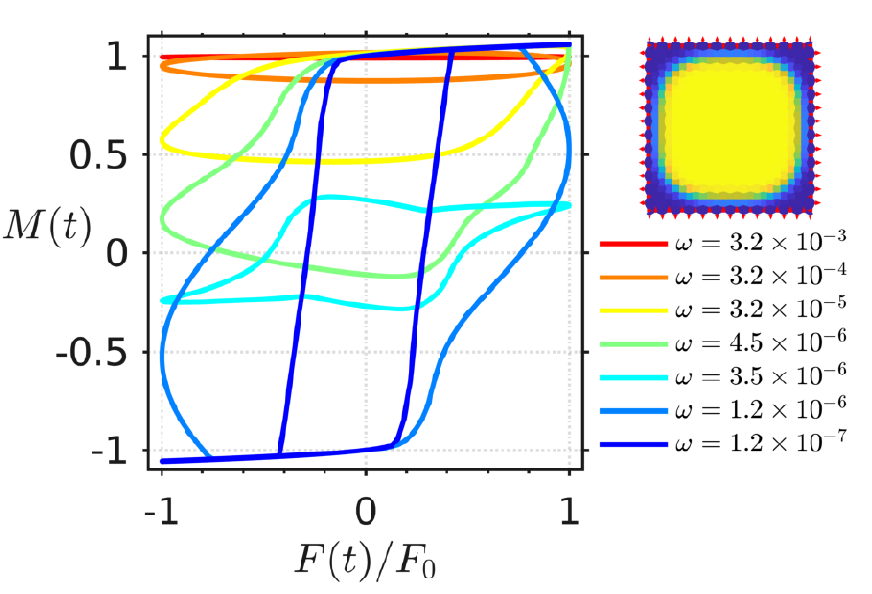}}
\end{subfigure}
\hfill
\caption{Frequency dependence of the hysteresis loops for compatible, cyclic forcing at the boundaries of the lattice, which tries to switch the entire metamaterial from $M =+1$ to $M=-1$ and back. Here $L_x = L_y = 25$ and the dimensionless interaction stiffness is $k_2 = 0.2$. The cycle proceeds counterclockwise along the loop. Inset: Snapshot of intermediate point on the hysteresis loop for $\omega = 1.2\times10^{-6} $ and the applied forces at the boundary nodes (red arrows). {\color{black} Videos of the cycles for $\omega = 3.2\times10^{-5}$, $4.5\times10^{-6}$, $3.5\times10^{-6}$, and $1.2 \times 10^{-6} $ are provided in Appendix F, Videos 1-4.}}
\end{figure}

\section{Dynamical Hysteresis}

We now investigate how this rich topological zoology affects dynamics. In square spin ice, monopoles and domain walls can be rearranged by external magnetic fields.
While an analogue of a bulk magnetic field acting on all unit cells at once is not readily available for mechanical systems, we can focus on the experimentally realizable case of forcing the boundaries of the mechanical metamaterial. First, we apply forces to the nodes of the cells at the boundary {\it in a compatible way}, i.e\ by alternating the sign of the force among consecutive nodes (see inset of Fig.~3) and modulating the intensity harmonically in time $F=F_0 \sin(\omega t)$, where $F_0$ is the force amplitude and $\omega$ is the dimensionless driving frequency. The characteristic relaxation time for the framework springs, $T_1 = \gamma/k_1 = 1$, where $\gamma$ is the linear drag coefficient, is taken as the unit of time (see Appendix A), and we study the dependence on the dimensionless driving frequency approaching the quasistatic limit, $\omega \ll 1$. 
When driven in such a manner, the system undergoes a hystersis cycle, with the shape of the hysteresis cycle governed by the dimensionless driving frequency $\omega$, see Fig.~3. When the intensity of the force $F(t)$ overcomes a coercive value $F_c$, the boundary cells all snap from $M_{\alpha \beta}=+1$ to $M_{\alpha \beta}=-1$, creating a domain wall loop sitting just inside the edge of the metamaterial. Once formed, this loop starts to contract to reduce its overall length and energy, see Fig.~3 inset. The subsequent internal dynamics depend on how the driving frequency $\omega$ compares to the speed of the internal domain wall contraction. 

At high $\omega$, the boundary force oscillates too rapidly for the system to mechanically respond, leading to a shallow penetration depth of stressed cells and leaving the bulk unchanged, with $M(t) \sim 1$ at all times. For less rapid forcing, edge cells invert, creating a circumferential domain wall. This domain wall begins to contract toward the center of the system, but it is subsequently annihilated by a second domain wall forming in the second half of the cycle, preventing it from penetrating into the bulk. Sample time series are shown in Appendix D,~Fig.~10, {\color{black} and corresponding videos are provided in Appendix F, Videos 1-4.} The fact that the second loop catches up with the first is evidence of an attractive interaction between the two loops, since otherwise each loop would follow the exact same trajectory, differing only by a time delay. Once the driving period $\tau = 2\pi/\omega $ is just large enough for the first domain wall to propagate inwards without being caught by the subsequent domain wall, a first-order dynamical transition appears at which the hysteresis loops change abruptly from being asymmetric, to being symmetric about $M(t)=0$. For the $k_2 = 0.2$ hysteresis loops shown in Fig.~3, the transition occurs as the driving frequency is reduced from $\omega = 4.5\times10^{-6}$ to $\omega = 3.5\times10^{-6}$. This behavior resembles the dynamic phase transition studied in kinetic Ising models driven at finite frequency by an external magnetic field~\cite{tome1990dynamic,rao1990magnetic,chakrabarti1999dynamic,berglund1999memory,korniss2000dynamic}, except that the transition we observe appears to be discontinuous.
{\color{black} We also note that it is natural to inject and manipulate domain walls via the boundaries of a mechanical metamaterial, whereas magnetic systems are typically driven via a bulk magnetic field.} In the quasistatic limit, $\omega \rightarrow 0$, the forcing period $\tau$ is far larger than the time needed for the internal loop to contract. Thus, the hysteresis loop is square.  
 
\begin{figure}[t!]
\begin{subfigure}[c]{\linewidth}
\centering
\resizebox{\linewidth}{!}{\includegraphics{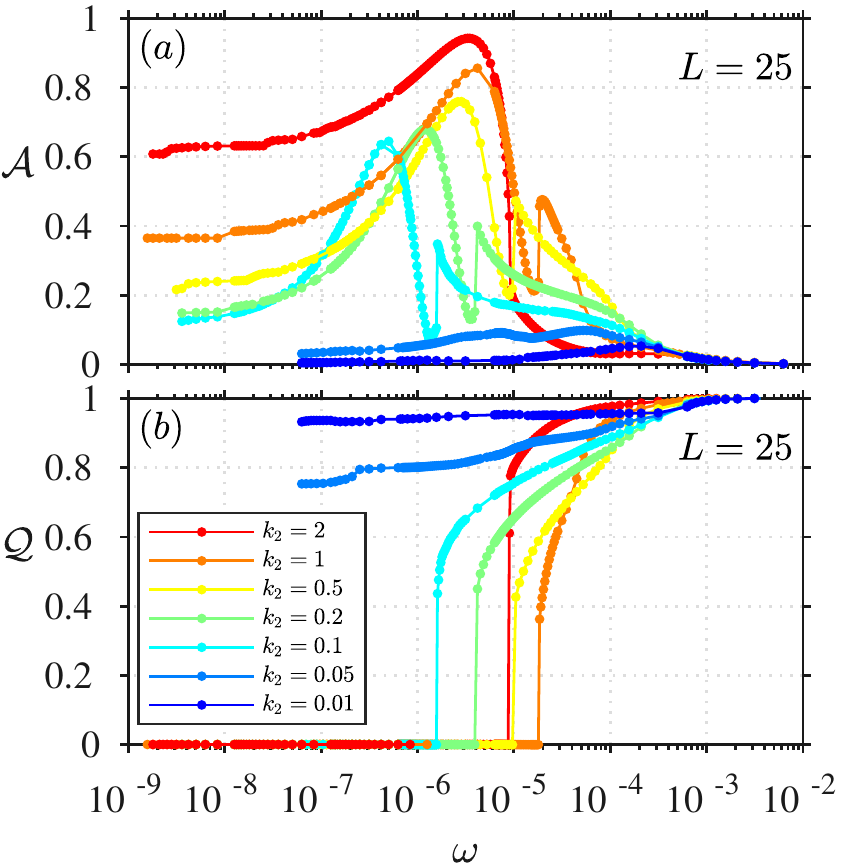}}
\end{subfigure}
\begin{subfigure}[c]{\linewidth}
\centering
\resizebox{\linewidth}{!}{\includegraphics{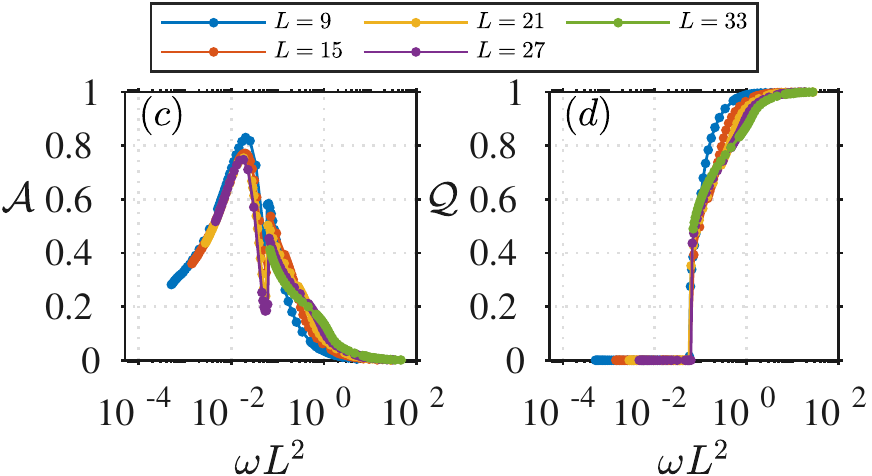}}
\end{subfigure}
\hfill
\caption{Frequency dependence of the normalized area $\mathcal{A}$ (a) and the time-averaged magnetization $\mathcal{Q}$ (b) for the compatible driving protocol, and compared for varying dimensionless interaction stiffness $k_2$. The characteristic time scales are increased by reducing the interaction strength $k_2$. (c)-(d) Finite-size scaling of $\mathcal{A}$ and $\mathcal{Q}$ for square lattices $L_x = L_y = L$ and $k_2 = 0.5$, confirming that the circular domain wall loops contract at a rate proportional to their curvature.}
\end{figure}
	
We quantify the hysteresis through its normalized area $\mathcal{A}(
\omega) = \frac{1}{4F_0} \int M dF $ and the time-averaged magnetization over one period $ \mathcal{Q}(\omega) = \frac{1}{\tau} \int M dt $ ~\cite{tome1990dynamic,rao1990magnetic,chakrabarti1999dynamic,berglund1999memory,korniss2000dynamic}. $Q=0$ when the cycle is symmetric, and $\mathcal{Q} \ne 0$ indicates an asymmetric cycle. In Fig.~4(a,b), we plot $\mathcal{A}$ and $\mathcal{Q}$ vs. $\omega$ for a set of $k_2$ values ranging from the binary, spin-ice limit to the elastic regime. For $k_2\geq0.1$, the dynamic transition from asymmetric to symmetric cycles is marked by a discontinuous jump in both $\mathcal{A}$ and $\mathcal{Q}$, with $\mathcal{Q}$ remaining zero below the {\it critical} frequency $\omega_c$.  $\omega_c$ can be related to $\tau_c$, the time it takes a circular domain wall to collapse, as $\omega_c=\pi/\tau_c$. When $\omega<\omega_c$ full inversion is achieved by the collapse of the circular domain wall before a new half cycle begins. When $\omega> \omega_c$ a new domain wall is created before the previous can collapse, and $M$ doesn't fully invert sign, leading to asymmetry. 

Another interesting frequency is the {\it resonant} frequency $\omega_r$, which corresponds to the maximum in the area $\mathcal{A}$. Here, just after reaching orientation $M=-1$, the system is driven to immediately return to orientation $M=+1$, once again. The maximum area, resonant frequency loop for $k_2 = 0.2$, $\omega_r = 1.28\times10^{-6}$ is shown in Fig.~3, and a time series of the collapse of the circular domain wall is shown in{ \color{black} Appendix D,~Fig.~10(m)-(p), and in Appendix F, Video 4}.

\begin{figure*}[t!]

\begin{subfigure}[c]{\linewidth}
  \includegraphics[width=1\textwidth]{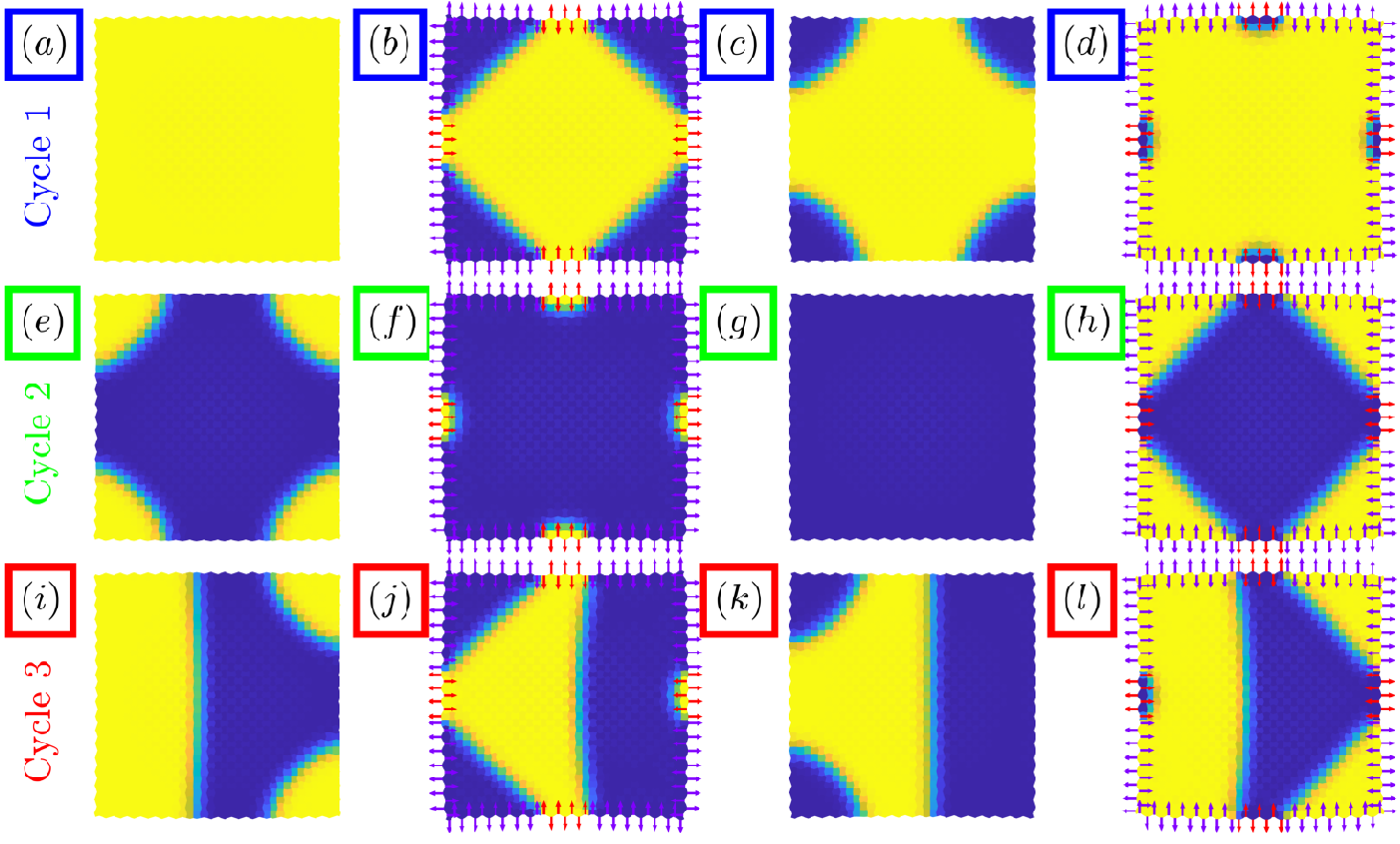} 
\end{subfigure}
\begin{subfigure}[c]{\linewidth}
  \includegraphics[width=1\textwidth]{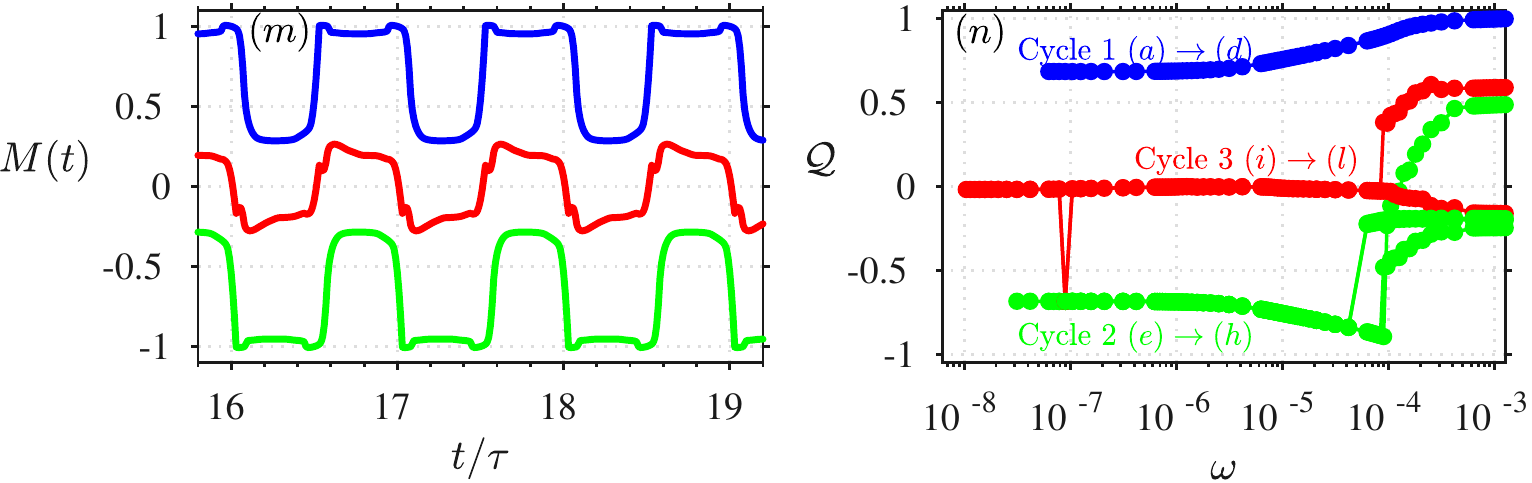} 
\end{subfigure}

\caption{Time series snapshots for three steady state cycles due to textured driving of the boundaries, with $L_x = L_y = 35$ and $k_2 = 0.5$. Four central segments (red arrows) are switched from $M_{\alpha \beta}=+1$ to $M_{\alpha \beta}=-1$, whereas four corner segments (purple arrows) switch from $M=-1$ to $M=+1$. Each column shows snapshots for a fixed phase of the textured forcing, with $\omega t = 0, \pi/2, \pi$, and $3\pi/2$ (the $M_{\alpha \beta}$ colorbar is the same as in Fig.~2). Starting from different initial metastable states, the metamaterial may be attracted to Cycle 1 (a)-(d), Cycle 2 (e)-(h), or Cycle 3 (i)-(l). (m) Magnetization $M(t)$ for the three cycles with driving frequency $\omega = 6.28 \times 10^{-7}$. Cycle 1 (blue) and Cycle 2 (green) are asymmetric $\mathcal{Q} = \pm 1 $ and related by a phase shift of $\pi$, while Cycle 3 (red) is symmetric $\mathcal{Q}=0$. (n) Frequency dependence of $\mathcal{Q}$ for six different initial conditions, one of which reaches Cycle 1 (blue), three reach Cycle 2 (green), and two reach Cycle 3 as $\omega \rightarrow 0$.}
\end{figure*}

A simple scaling analysis clarifies the significance of the time scale $\tau_c$, which is the time scale which sets the values of both $\omega_c$ and $\omega_r$. $M$ is a nonconserved order parameter undergoing dissipative dynamics, so we expect model-A dynamics to describe the domain wall motion~\cite{chaikin1995principles}. For a circular domain wall, it has been shown that the circle contracts at a rate proportional to its mean curvature: $ \frac{dR}{dt} \propto \frac{1}{R}$~\cite{cahn1977microscopic,allen1979microscopic}. Consequently, the rate of change of the area inside the circle $ v = \frac{d(\pi R^2)}{dt} \propto R\frac{dR}{dt} $ is constant. The total time for the circle to contract on itself is then approximately $\tau_c \propto L^2 / v$. In Fig 4(c)-(d), we confirm this time dependence by performing finite-size scaling of the hysteresis loops for square lattices of linear sizes $L=9 - 33$. The hysteresis loop areas $\mathcal{A}$ and time-averaged magnetization $\mathcal{Q}$ collapse when plotted against the rescaled time $ \omega L^2$, confirming that the circular domain walls contract at a speed proportional to their curvature.

\section{Textured Forcing Protocol}
	
We have demonstrated how the domain wall dynamics control the probe response of the system in the very simple case of a compatible drive. However, our system allows us to ``inject'' desired domain walls for different dynamical responses, by finely controlling the detail of the applied force pattern. For instance, wherever we skip the alternation of sign of the applied force along the boundary we inject a domain wall. This can only be done an even number of times, and each will correspond to a domain wall ending at the interface.

In the compatible forcing protocol, only one static metastable configuration matches the forcing texture at maximum (minimum) amplitude of the textured boundary forces, at times such that $\omega t = \pi/2$ ($\omega t = 3\pi/2$), namely, the global ground state $M=+1$ ($M=-1$). Many such matching, metastable states are made possible for an applied forcing texture composed of alternating $M_{\alpha \beta}=\pm1$ segments (see boundary arrows in Fig.~5). In Appendix E, we demonstrate that the number of these states is given by the Catalan numbers~\cite{wolfram}, which grow nearly exponentially as a function of the number of alternating force segments. Below, we demonstrate that different initial conditions with different internal domain wall configurations may be attracted to topologically-distinct steady cycles. These cycles are always between a subset of the static metastable configurations (and their inverses) for which the boundary configurations and the arrangement of internal domain walls match compatibly with the forcing texture at its extremum values.

First, we study a relatively simple textured forcing protocol. {\color{black}Figure~5} illustrates three steady cycles which appear when the system is driven with $8$ alternating forcing segments. Metastable states which match this forcing texture must contain $4$ internal domain walls. There are $14$ such static configurations: three of these distinct states appear in the second column of Fig.~5, {\color{black}panels $(b), (f) $ and $(j)$, while their inverses appear in the last column: (h) is the inverse of (b), (d) is the inverse of (f), and (l) is the inverse of (j) rotated by $\pi$. There are in fact $6$ states with nondegenerate domain wall configurations, shown in Appendix E, Fig.~11, and all $14$ topologically-distinct states may be generated by applying $\pi/2$ rotations to these $6$ states.}

We tested a large number of different initial conditions: for each initial condition we imposed $8$ alternating $M_{\alpha \beta}=\pm1$ segments as fixed boundary conditions, thus giving one of the $14$ topologically-distinct metastable configurations as an initial condition. At high frequencies, the domain wall penetration depth is shallow, and a variety of different steady states appear, reflecting the different internal domain wall configurations of the initial conditions. These can be seen by the different $Q$ values at large $\omega$ in {\color{black}Fig.~5(n)}. Interestingly, in the quasistatic limit, the internal domain wall configurations are erased, yet the system tends to one of three steady-state $M(t)$ times series, each giving a signature $Q$ value. Figure~5(m) shows these three distinctive magnetization time series. The panels in each column show the magnetization configurations $M_{\alpha \beta}$ at times $\omega t = 0, \pi/2, \pi,$ and $3\pi/2$ in the steady state, when the textured forcing amplitude is zero or extremal.

For Cycle 1, $M(t)$ varies asymmetrically between $0.3$ and $1$, with time-averaged magnetization $\mathcal{Q}=0.6$. In Cycle 2, $M(t)$ varies between $-0.3$ and $-1$, and $\mathcal{Q}=-0.6$. This cycle mirrors Cycle 1, but with a phase shift of $\pi$. This phase shift can be understood by comparing the second and last columns of Fig.~5, when the force is at it maximum (minimum) value. There are two distinct topological configurations that match the boundary conditions, one with domain walls spanning the four corners [Fig.~5(b) or Fig.~5(h)], and another with semicircular loops on each side [Fig.~5(d) or Fig.~5(f)], and upon reversing the sign of the boundary force these two configurations transform into one another. Finally, Cycle 3 is symmetric, with $M(t)$ varying between $\pm0.3$, giving $\mathcal{Q} = 0$. This cycle alternates between inverse variants of a configuration with one long domain wall spanning the center and two corner spanning domain walls [Fig.~5(j) or Fig.~5(l)]. Cycle 3 does not have a phase-shifted counterpart. Rotations of the configuration in Fig.~5(j) give four distinct domain wall arrangements, so we consider Cycle 3 to have four variants, giving a total of $6$ distinct steady state cycles arising for this simple textured forcing protocol.

For the compatible drive studied in Sec. IV, starting from an initial condition with complicated internal domain wall configurations will not lead to multiple cycles. When compatibly cycling initial states containing several domain walls, domain walls within the starting configurations are always annealed away after some transient number of cycles. Eventually, the system always returns to the same hysteresis loops of the previous section, and all memory of the starting state is erased. The duration of long-lived transients grows longer for larger $\omega$. Similarly, if the $8$ segment driving protocol of Fig.~5 were applied to a random configuration containing some complex domain wall pattern not necessarily compatible with the applied force, one of the $6$ distinct steady states cycles will eventually be reached.

\begin{figure*}[t!]
\centering 
\begin{subfigure}[c]{\linewidth}
    \includegraphics[width=1\textwidth]{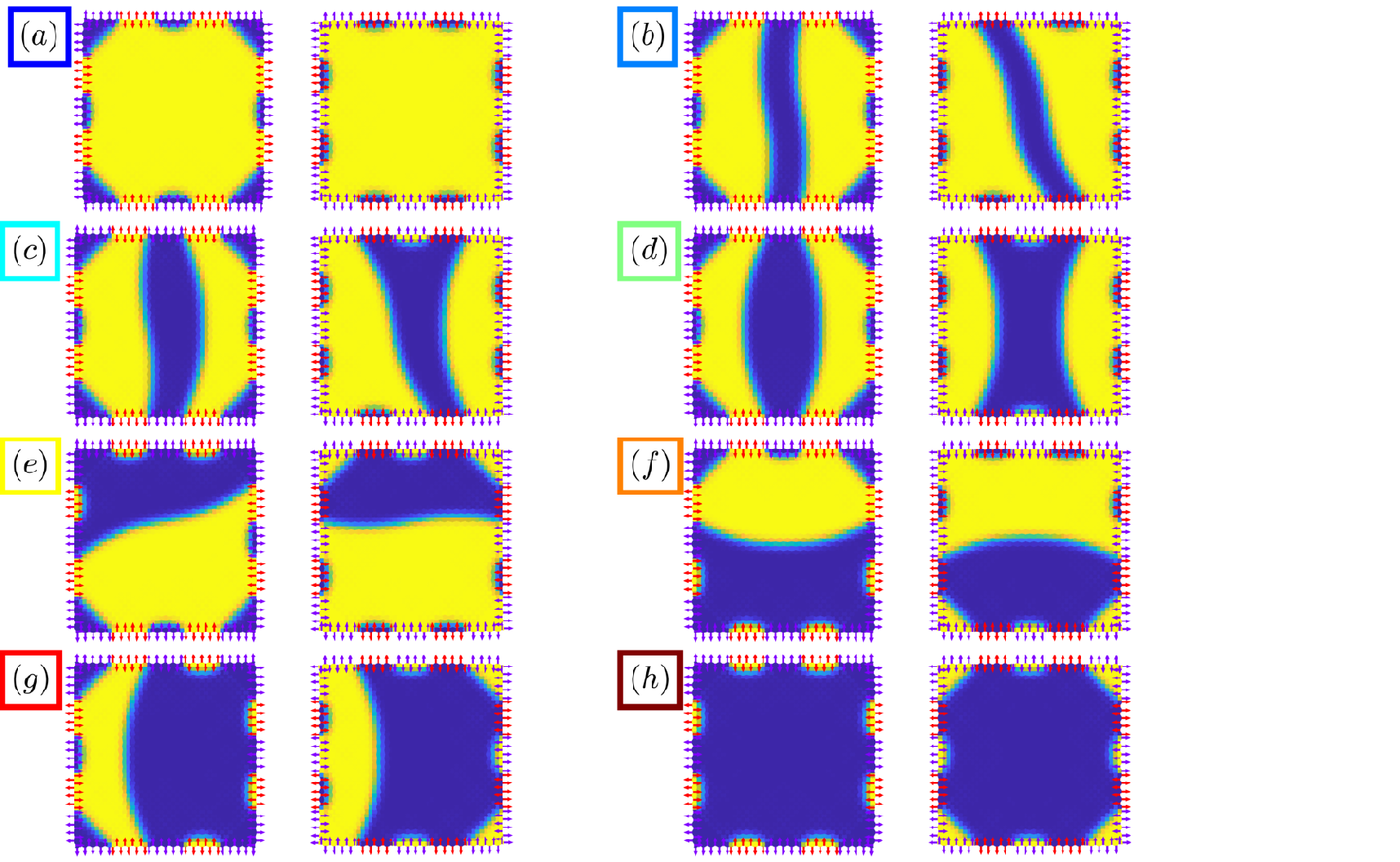}
\end{subfigure}
\begin{subfigure}[c]{\linewidth}
  \includegraphics[width=1\textwidth]{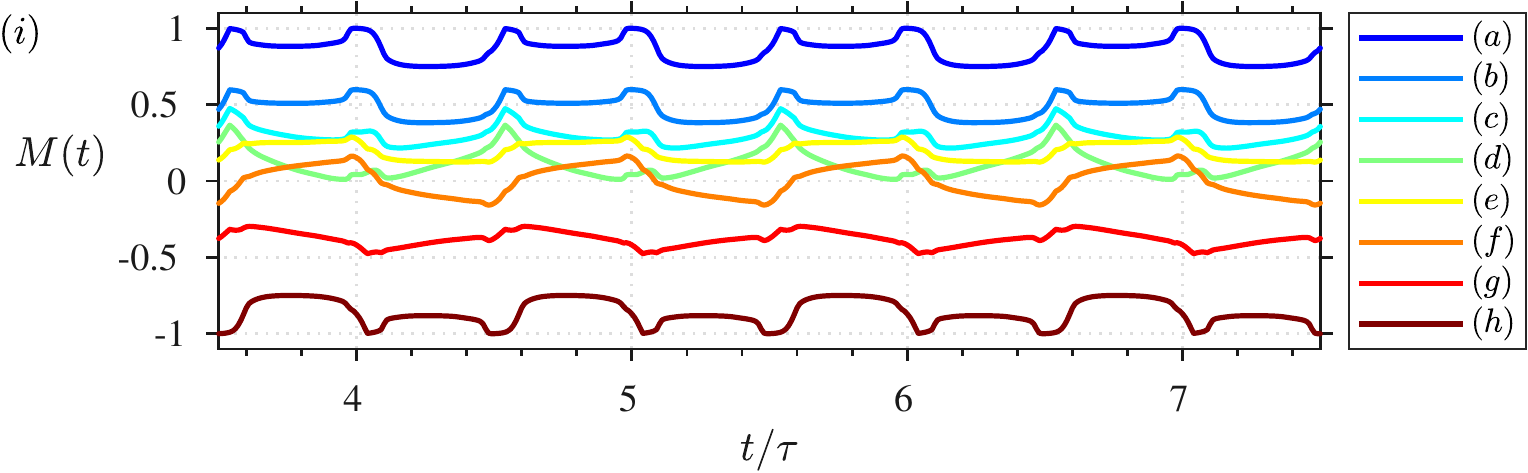} 
\end{subfigure}

\caption{(a)-(h) Topologically-distinct domain wall configurations which occur at the maximum and minimum values of the textured forcing, $\omega t = \pi/2, 3\pi/2$, with sixteen alternating $M_{\alpha \beta}=\pm1$ textured forcing segements along the boundary (red and purple arrows). $L_x=L_y=45$ and $k_2 = 0.5$, and the cycles were found for fixed small frequency $\omega = 1.9\times10^{-6}$ (the $M_{\alpha \beta}$ colorbar is the same as in Fig.~2). (i) Magnetization time series corresponding to the snapshots shown above illustrating the rich variety of steady state hysteresis cycles arising from this textured forcing protocol.}
\label{fig6}
\end{figure*}

The ultimate cycle a given initial condition is attracted to depends critically on the initial domain wall topology: for example, an initial condition needs to have at least one domain wall spanning the width of the metamaterial to access Cycle 3. However, the initial topology does not guarantee the resulting cycle that the state will be attracted to, and finer details of the internal domain wall curvatures and small variations in the internal configuration determine the cycle which a particular initial arrangement is attracted to. For example, we found that different initial conditions each with two long domain walls [see Appendix E, Fig.~11(d)] sometimes reach Cycle 3, and sometimes reach Cycle 1 or Cycle 2. Finer details of the internal domain wall arrangements can also affect the stability of the distinct cycles. We note that Cycle 3 appears to be less stable: for some initial conditions, it appears to be the steady state over a wide range of frequencies, but then an abrupt switch may occur to Cycle 1 or Cycle 2. Except for the anomaly at $\omega\sim10^{-7}$, examples shown in {\color{black} Fig.~5(n)} (red curves) prefer Cycle 3 at least down to $\omega \sim 10^{-8}$. {\color{black} Sample videos of Cycle 1 and Cycle 3 are given in Appendix F, Videos 5 and 6.}

Finally, we demonstrate that the variety of the distinct hysteresis cycles can be easily increased by adding more alternating forcing segements along the boundary. Doubling the number of boundary defects, we tested a forcing protocol with $16$ alternating $M_{\alpha \beta}=\pm1$ segments along the boundary. Compatible initial conditions for this forcing protocol must have $8$ internal domain walls, and there are a total of $1,430$ distinct ways to place the domain walls (see Appendix E). Among these, each state has some trivial degeneracy of $1$, $2$, $4$, or $8$ from rotations or reflections (see Appendix E, Fig.~12). A simple lower bound may be attained by assuming the maximum degeneracy for each state, implying that there are at least $1,430/8 \approx 178 $ different domain wall configurations. 

{\color{black} By fixing the boundary conditions and quenching the energy as in Sec. III, we manually identified $63$ distinct states out of a sample set of $2,000$ trial cases, and we expect additional rare states would be found for large ensembles or could be directly imposed on the metamaterial. Snapshots of these energetically distinct domain wall configurations are given in Appendix E, Fig.~13 and Fig.~14.} Taking these states as initial conditions, and driving them at low frequency $\omega = 1.9\times10^{-6}$, we observe many distinct steady state cycles, $8$ examples of which are shown in Figure~6. Two panels for each cycle show the topologically-distinct configurations which appear at the maximum and minimum textured forcing amplitude. Some of the cycles, like those shown in (a), (f) and (g), are relatively common and reached by many of the trial initial conditions. Others, like (b), (c), (d), and (e) are more rare. {\color{black} Sample videos of the cycles corresponding to Fig.~6(a-d) are provided in Appendix F, Videos 7-10.} Additional cycles occurred which are rotational and phase-shifted counterparts to these example cycles, and we expect to find more cycles by testing a larger ensemble of initial configurations. 

\section{Discussion}

We have demonstrated that our mechanical analogue of spin-ice exhibits real-space topological defects, similarly to magnetic and other realizations of ASI. These domain walls tend to escape to the system boundaries, but they can sometimes be immobilized depending on the ratios of spring stiffnesses used to construct the metamaterial or on imposed boundary conditions. As the dimensionless interaction stiffness $k_2$ is reduced, the domain walls become slower and more curved, until their motion ceases {\color{black} in the ice-like limit $k_2 \ll k_1$}. Uniform inversion of the boundaries creates circular domain wall loops which contract upon themselves, setting up a dynamic hysteresis cycle between the competing ground state configurations. There appears to be a first-order dynamic phase transition associated with this uniform driving protocol. The system exhibits a novel dynamic complexity when textured forcing is applied at the boundary: different initial configurations of the network are attracted to multiple, topologically-distinct steady cycles. The number of cycles is closely related to the multiplicity of topologically-distinct static states consistent with the texture of the boundary forcing. Further, the possibility of many distinct ways of connecting domain walls inside the material protects these cycles: once a steady cycle is achieved, the system cannot easily be pushed into a new cycle because of the large energy barriers which would need to be crossed in order to change the positions of domain walls from their current arrangement.

This potential for many steady cycles arising for a single driving texture relies on the ability to manipulate the boundary of the metamaterial, unlike ASI and other systems which are {\color{black} typically} driven by bulk fields. Thus, our mechanical spin-ice system opens up further exploration of rich dynamical phenomenology which have no counterparts in magnetic systems. Our work opens a new direction of using periodic prestressed mechanical metamaterials in order to generate complex energy landscapes, which in turn lead to multiple steady states. {\color{black} It should be fruitful to explore additional driving protocols, as well as modified lattice geometries. Additional rich hysteresis behavior could be generated by driving opposite pairs of the boundaries out of phase from one another, or driving different sides with different amplitudes. Specific domain wall configurations could be imposed by programmed protocols that manipulate parts of the boundary one by one in a specific sequence. Moreover, the degeneracies of distinct metastable configurations inside the material can be engineered by changing the shape of the boundary or the spacing of forcing defects along the boundary, possibly allowing for more rich behaviors. Finally, it should be interesting to consider more elaborate spin-ice mechanical metamaterials, but based on lattices that are inherently frustrated and thus have extensively degenerate ground state configurations~\cite{gilbert2014emergent,lao2018classical}. Studying such lattices could promote the understanding of the complex dynamics and memory of amorphous solids and other glassy systems. As such, it will be useful to study the defects and possible dynamic hysteresis in such generalized mechanical spin-ice systems.}

\begin{acknowledgments}
    We thank Martin van Hecke, Yoav Lahini, Erdal O\u{g}uz, and Ben Pisanty for helpful discussions. This work was partially funded by the Israel Science Foundation Grant No. 968/16 and the National Science Foundation Grant No. NSF PHY-1748958. YS thanks the Center for Nonlinear Studies at Los Alamos National Laboratory for its hospitality. The work of CN was carried out under the auspices of the U.S. Department of Energy through the Los Alamos National Laboratory, operated by Triad National Security, LLC (Contract No. 892333218NCA000001). CM thanks Fulbright and the United-States Israel Educational Foundation for financial support.
\end{acknowledgments}

\appendix

\section{Computational Methods}

Our mechanical metamaterial is modeled as an elastic spring network simulated using overdamped dynamics or the method of steepest energy descent. For a network of nodes with equal mass $m$, the overdamped equation of motion for a given node with position $\vec{r}_i$ is \begin{equation} \frac{d\vec{r}_i}{dt} = \frac{1}{\gamma}\sum_{\langle ij \rangle} -k_{ij}(e_{ij}-l_{ij})\hat{r}_{ij}, \end{equation} where the spring connecting node $i$ to node $j$ has spring constant $k_{ij}$, relaxed length $l_{ij}$, and current extension $e_{ij}=|\vec{r}_i - \vec{r}_j|$, and $\gamma$ is the linear drag damping force coefficient. We adopt units such that relaxation time scale $T_1 = \gamma/k_1 = 1$, and assume the system has an overdamped response such that the viscous damping time scale is much faster than the unit time $ m/\gamma \ll \gamma/k_1 = 1$. In the characteristic relaxation rate for interaction springs is ${T_2}^{-1} = k_2$ and for the compressive springs is $ {T_3}^{-1} = k_3$. Thus, the fastest relaxation rate in the system is given by ${k_3}^{-1} = 0.2$. We integrated the equations of motion with variable time step $\Delta t = 0.01-0.1$. For characterizing the various metastable conditions in Sec. III, we use free boundary conditions, with no constraints or additional forces on the nodes. In Sec. IV and Sec. V, we drive the lattice from the boundary by applying additional forces to the boundary nodes, perpendicular to the boundary, with magnitude $|F_i|=F_0\sin(\omega t)$, and alternating sign as determined by the desired driving texture, taking $F_0=1.5$ throughout.

\section{Spin Ice Vertex Classification}

\begin{figure}[t!]
\begin{subfigure}{\linewidth}
\resizebox{\linewidth}{!}{\includegraphics{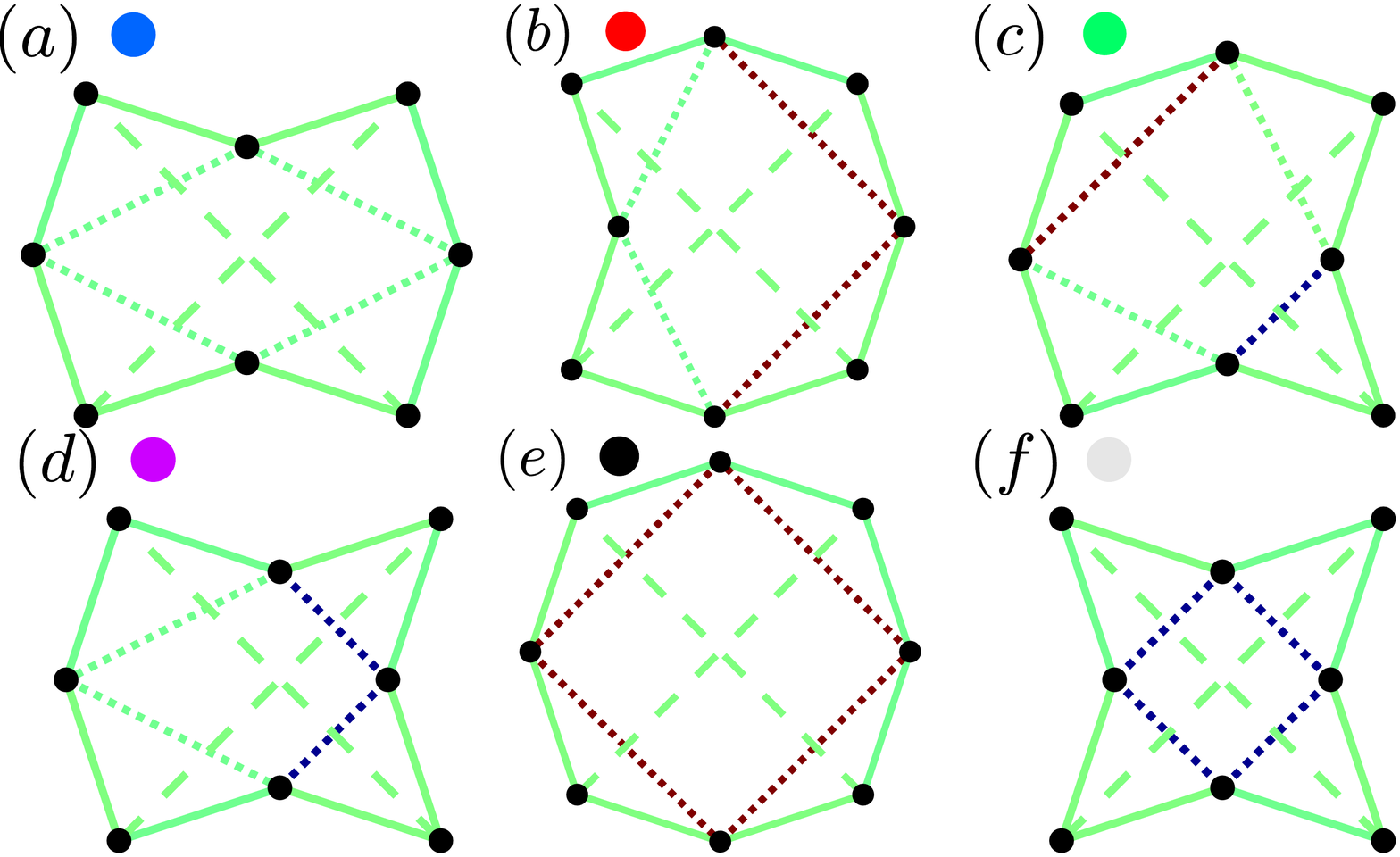}}
\end{subfigure}
\centering
\caption{(a)-(f) Idealized vertex configurations for the square unit cell showing the six possible vertex types in order of increasing energy. Solid lines show the framework springs, dotted lines show the interaction springs, and dashed lines show the diagonal compressive springs. Relaxed springs are colored green, compressed springs are colored blue, and springs under tension are colored red. See strain color map in Fig.~8.}
\end{figure}

Within ASI systems, spin configurations at a vertex are classified by their energy and their topological charge $q$, which is defined as the number of spins pointing into the vertex minus the number of spins pointing away from the vertex. Similarly, in the limit that the interaction springs are much weaker than the framework springs, and both are much weaker than the compressive springs, $k_2 \ll 1 < k_3$, we can obtain an approximate vertex hierarchy in energies for the square unit cells by treating the four faces of the unit cell as pseudo-spins. In this limit, we require that the two diagonal compressive springs and the eight framework springs must all be fully relaxed. This means that each of the four face nodes of the unit cell must bend inwards or outwards by an amount $\pm \delta = l_1 \sin(\theta^*) $. As discussed in the main text, all of the springs are relaxed when two opposing faces of the unit cell bend inwards and the remaining two faces bend outwards, and the total elastic energy is zero. Excitations happen when two adjacent faces both point outwards or both point inwards, causing the interaction spring connecting the two faces to be stretched or compressed relative to its relaxed length $l_2 = \sqrt{2}l_1$. Figure~7. shows the strain $\epsilon$ on all springs for all six possible vertex configurations. Unit cell configurations very similar to these idealized vertex configurations appear within the bulk of the metamaterial, as shown in Fig.~8. Due to the unit-cell geometry, the energetic cost of compressing an interaction spring is slightly larger than the cost of stretching the same interaction spring. For a given value of $\theta^* = \cos^{-1}{(l_3/2\sqrt{2})}$, the energy of a single interaction spring is given by $ E/k_2 = 0.5 \Delta l ^2 = (\sqrt{1 \pm \sin{2\theta^*}} - 1)^2$, where the plus sign is for a stretched interaction spring and the minus sign for a compressed spring. Simulations in the main text were carried out using $\theta^*=18.2^{\circ}$. Adding up the energy contribution from each stressed interaction spring, we find the following energy hierarchy for the vertices: $E_{a}/k_2 = 0$, $E_{b}/k_2 = 0.1376$, $E_{c}/k_2 = 0.2$, $E_{d}/k_2 = 0.2626$, $E_{e}/k_2 = 0.2752$, and $E_{f}/k_2 = 0.5252.$ We may also assign the topological charges $q_{a} = 0$, $q_{b} = -2$, $q_{c} = 0$, $q_{d} = + 2$, $q_{e} = -4$, and $q_{f} = +4$. Note that in magnetic ASI, the energy for the $q=\pm2$ and $q=\pm4$ monopoles is independent of the sign of the charge $q$, whereas this degeneracy is lifted due to the asymmetry between stretching and compressing the interaction springs in our square mechanical ice.

\begin{figure}[t!]
\begin{subfigure}{\linewidth}
\resizebox{\linewidth}{!}{\includegraphics{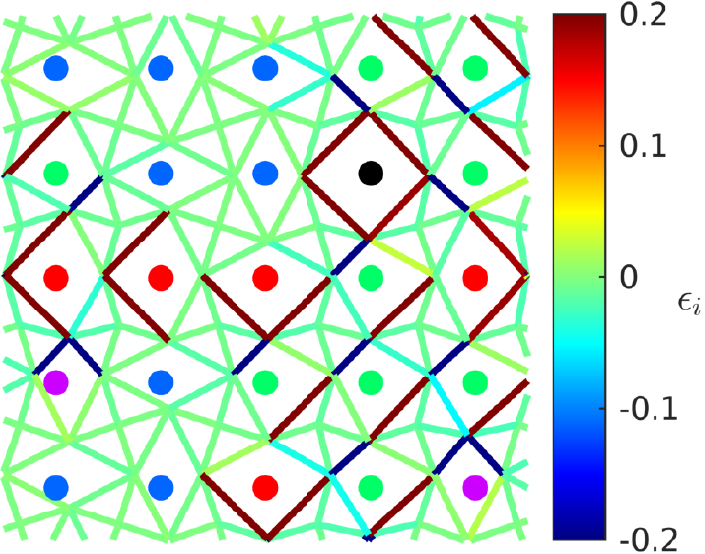}}
\end{subfigure}
\centering
\caption{Zoom on a small region for a metamaterial with $k_2 = 0.1$ showing strains $\epsilon_i$ for the framework and interaction springs. Unit cells which closely approximate the ideal vertex types are marked with colored dots according to the color scheme defined in Fig.~7. For clarity, the compressive springs are not shown.}
\end{figure}

\section{Domain Wall Speed}

\begin{figure}[t!]
\begin{subfigure}{\linewidth}
\resizebox{\linewidth}{!}{\includegraphics{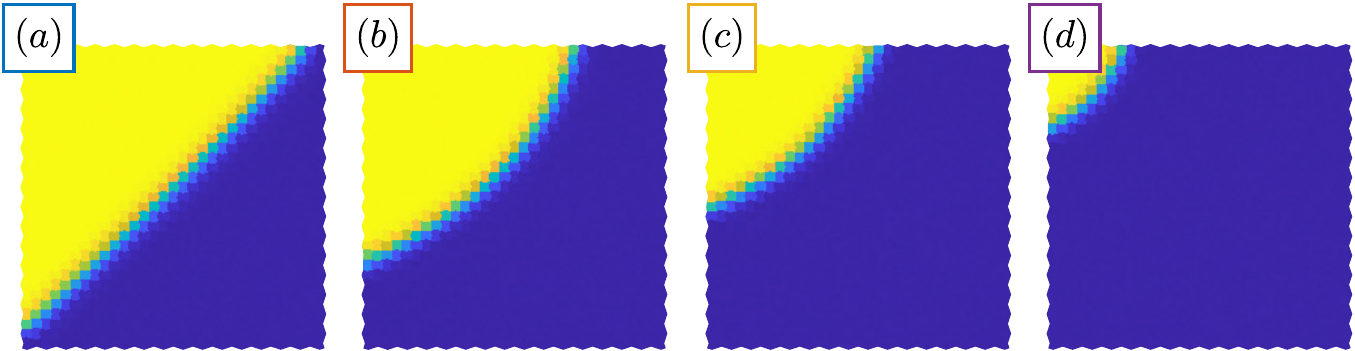}}
\end{subfigure}

\begin{subfigure}{\linewidth}
\resizebox{\linewidth}{!}{\includegraphics{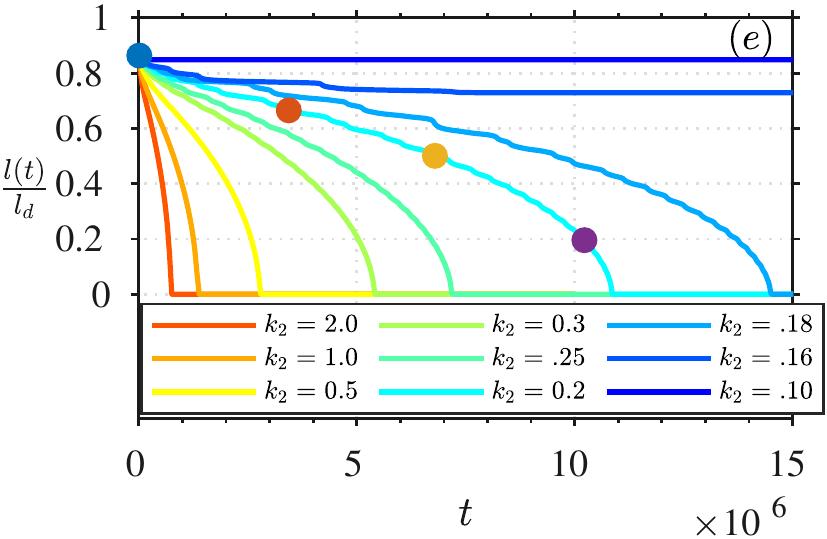}}
\end{subfigure}
\caption{(a)-(d) Time series of single corner-spanning domain wall length as it is expelled from the metamaterial for interaction spring stiffness $k_2 = 0.2$. (e) Measurements of the rate of decay of the total $M_{\alpha \beta}=0$ contour line length marking the moving domain wall. Colored dots correspond to panel labels (a)-(d). For interaction spring stiffness $k_2 < 0.18$ the domain wall gets stuck in place and remains forever.
}
\end{figure}

In order to quantify the domain wall speed, we compare the gradient descent dynamics for configurations starting with a long corner-spanning domain wall. We use a system of size $ L_x = L_y = 31 $, and prepare the initial configuration with a long domain wall spanning two adjacent sides and starting just above the diagonal of the metamaterial. During the dynamics, the domain wall decreases its length by escaping towards the corner of the metamaterial. In Fig.~9., we plot the time dependence of the contour length $l(t)/l_d$, where the diagonal length is $l_d = 2 \sqrt{2} L_x$, of the $M_{\alpha \beta} = 0$ contours for decreasing dimensionless interaction stiffness $k_2$. The escape time grows larger for decreasing $k_2$, until below a value $k_2\approx0.18$, the initial corner-spanning domain wall remains in place forever without escaping to the edge. These results are consistent with the change in the nature of the hysteresis cycles seen in Fig.~4, where for $k_2 = 0.01-0.05$, the hysteresis cycle is always restricted to perturbations along the boundaries even in the quasistatic limit $\omega \rightarrow 0$.

\section{Time Series for Simple Boundary Driving Protocol}

Fig.~10. illustrates several time series of the magnetization $M_{\alpha \beta}$ for the hysteresis cycles shown in Fig.~3. The first four panels, (a)-(d), show the trivial high frequency behavior for which the domain wall never penetrates into the bulk of the metamaterial. For a frequency slightly larger than the critical frequency $\omega_c$, panels (e)-(h), the domain wall loop created on the second half of the cycle catches up and annihilates with the first domain wall loop, creating a quadrupolar pattern. Unit cells deep within the center of the lattice are never flipped, keeping their magnetization value $M_{\alpha \beta} = +1$. In contrast, for a frequency just below the critical frequency, panels (i)-(l), the two circular loops move concentrically without colliding. At the resonant frequency, $\omega_r$, panels (m)-(p), the first circular loop contracts completely just as the second circular loop is nucleated at the boundary of the lattice. Finally, in the quasistatic limit $\omega \rightarrow 0$ (not shown), the loops contract rapidly compared to the driving period $\tau$, and so the metamaterial spends most of the time waiting in either state $M=+1$ or $M=-1$.

\section{Multiplicity of Topologically Distinct Configurations}

\begin{figure}[b!]
\begin{subfigure}{\linewidth}
\resizebox{\linewidth}{!}{\includegraphics{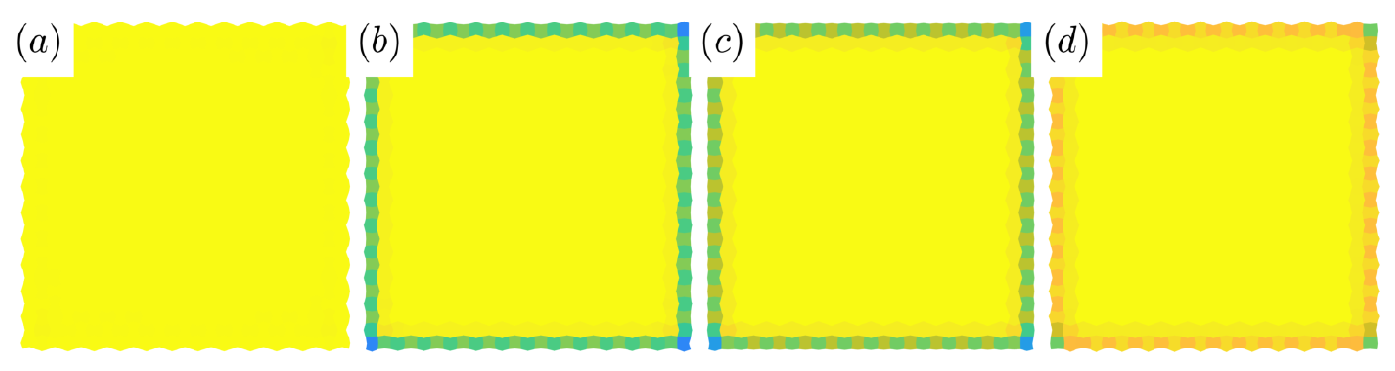}}
\end{subfigure}
\begin{subfigure}{\linewidth}
\resizebox{\linewidth}{!}{\includegraphics{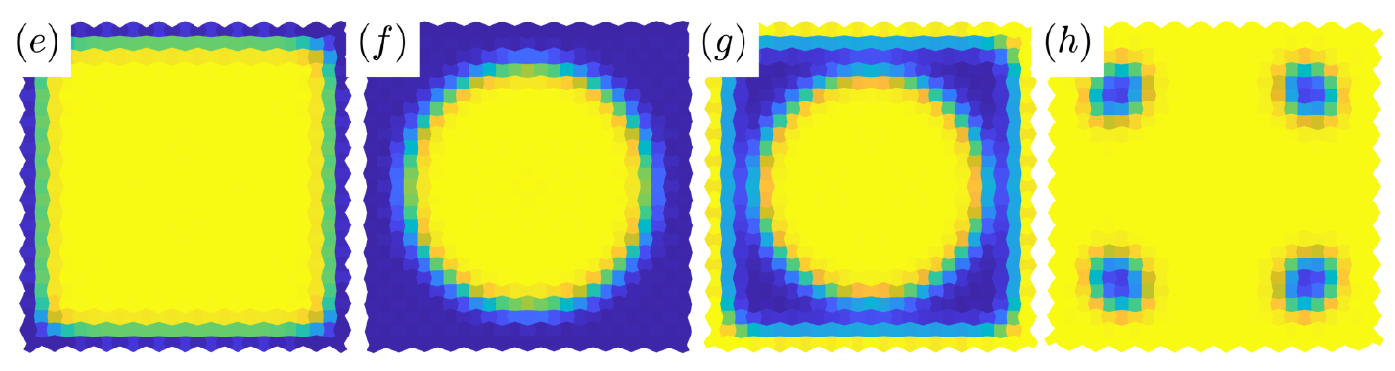}}
\end{subfigure}
\begin{subfigure}{\linewidth}
\resizebox{\linewidth}{!}{\includegraphics{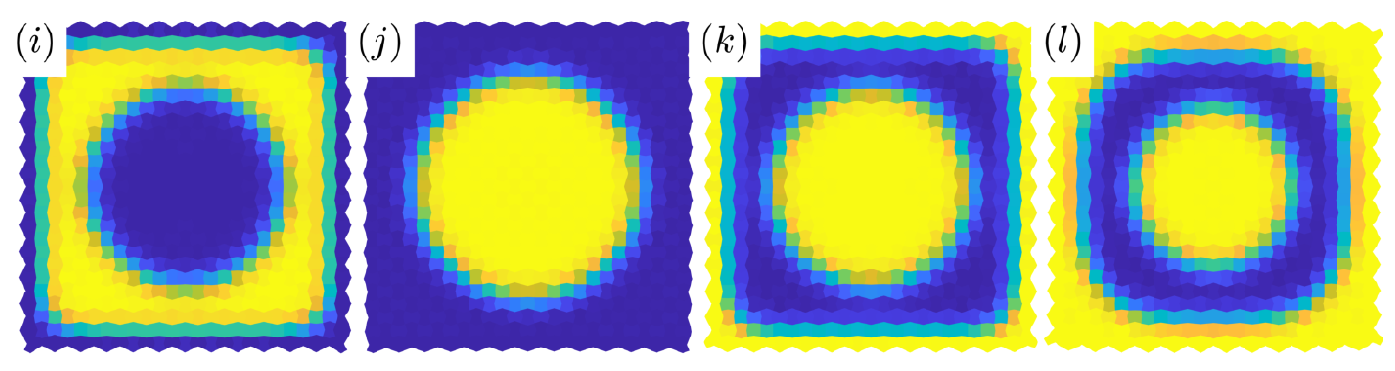}}
\end{subfigure}
\begin{subfigure}{\linewidth}
\resizebox{\linewidth}{!}{\includegraphics{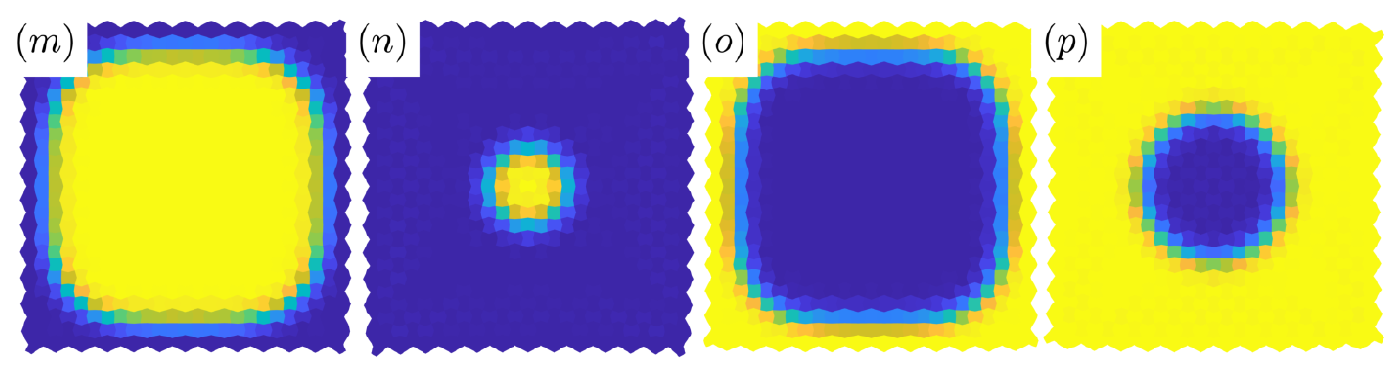}}
\end{subfigure}
\caption{Steady state time series examples for the simple, uniform driving protocol for lattices of size $L_x=L_y=25$ and interaction spring stiffness $k_2 = 0.2$. Panels (a)-(d) show snapshots over one period for the trivial behavior at high frequency, $\omega = 3.14\times10^{-4}$ where only boundary cells are deformed. Panels (e)-(h) illustrate the cycle at driving frequency $\omega = 4.48\times10^{-6}$, just above the critical frequency $\omega_c$. Panels (i)-(l) show the change in behavior at $\omega = 3.49\times10^{-6}$, just below $\omega_c$. Finally, the behavior at the resonant frequency $\omega_r = 1.28\times10^{-6}$ is shown in panels (m)-(p).}
\end{figure}

\begin{figure}[t!]
\begin{subfigure}{\linewidth}
\resizebox{\linewidth}{!}{\includegraphics{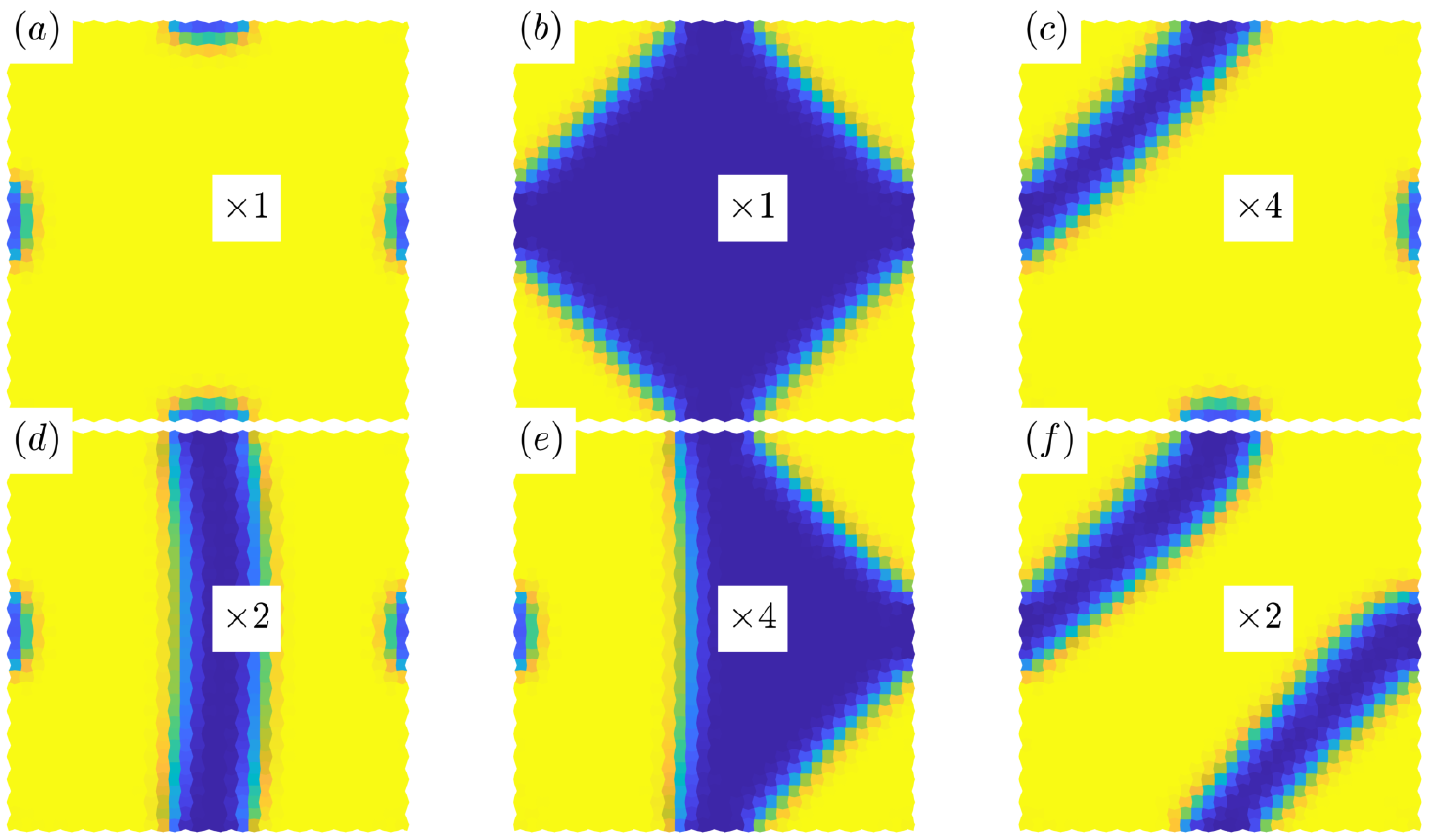}}
\end{subfigure}
\centering
\begin{subfigure}{0.8\linewidth}
\resizebox{\linewidth}{!}{\includegraphics{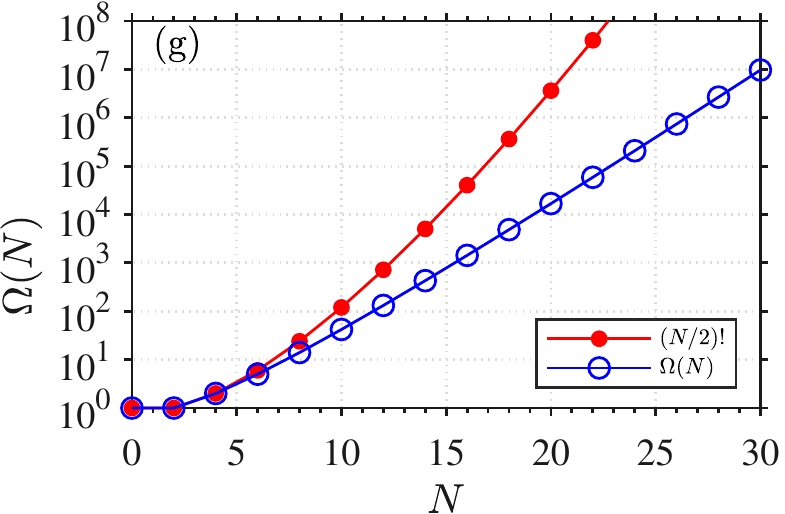}}
\end{subfigure}

\hfill
\caption{Panels (a)-(f) illustrate the six energetically distinct states possible for $N$=8 $A$ and $B$ segments along the boundary. Adding up the labeled rotational degeneracies of each state ($\times1,\times2, \times4$) gives the predicted number of distinct configurations: $\Omega(8) = C_{4} = 14$. (g) Exact enumeration of the total possible number of topologically distinct arrangements as a function of $N$ (blue), as well as their upper limit (red).}
\end{figure}

\begin{figure}[b!]
\begin{subfigure}{\linewidth}
\resizebox{\linewidth}{!}{\includegraphics{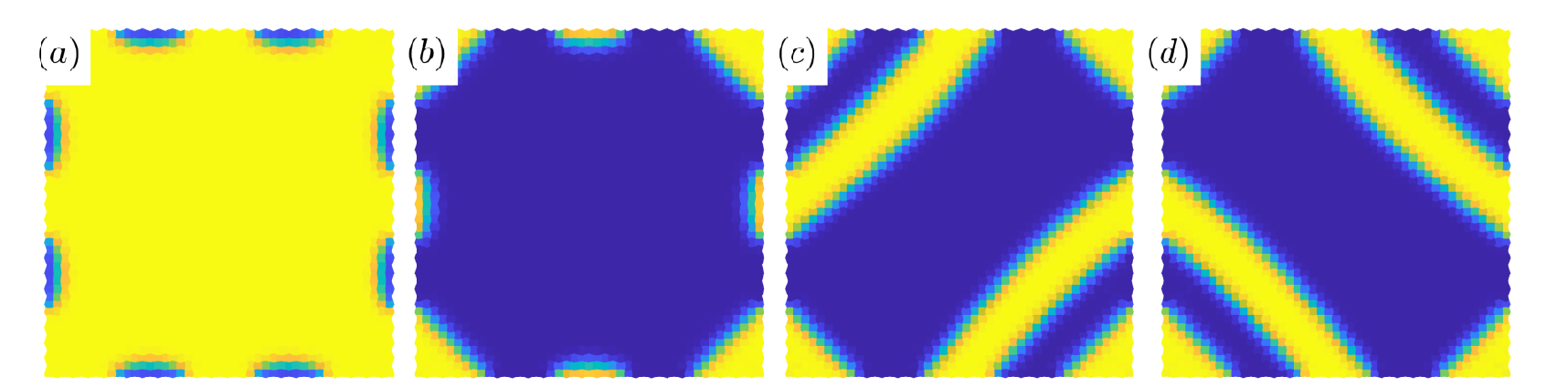}}
\end{subfigure}
\begin{subfigure}{\linewidth}
\resizebox{\linewidth}{!}{\includegraphics{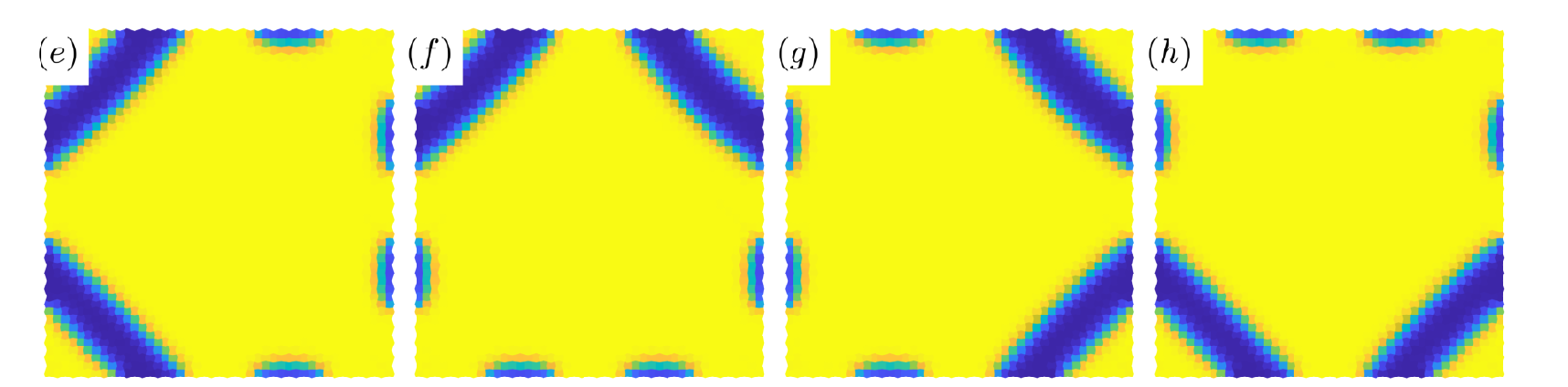}}
\end{subfigure}
\begin{subfigure}{\linewidth}
\resizebox{\linewidth}{!}{\includegraphics{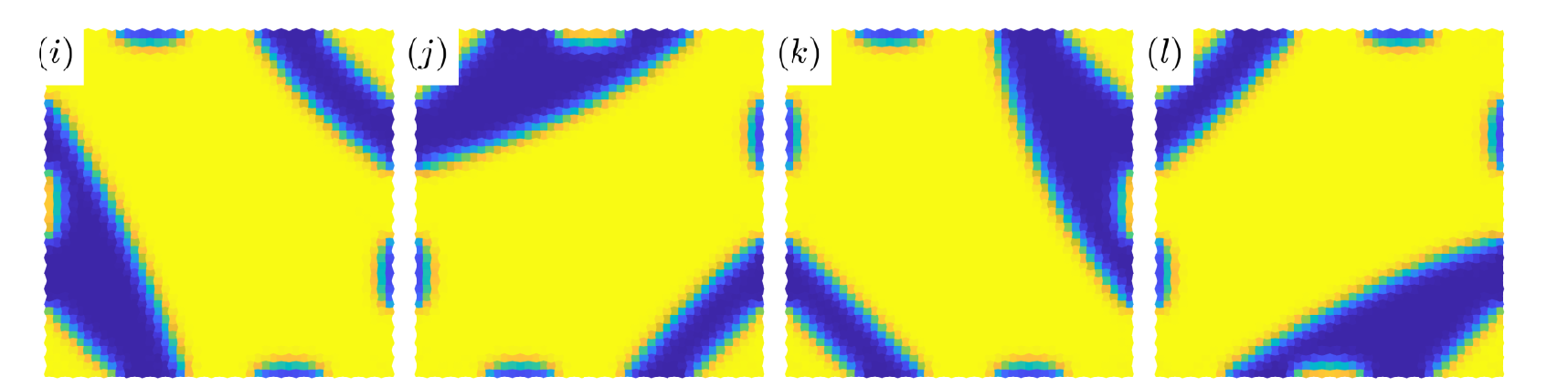}}
\end{subfigure}
\begin{subfigure}{\linewidth}
\resizebox{\linewidth}{!}{\includegraphics{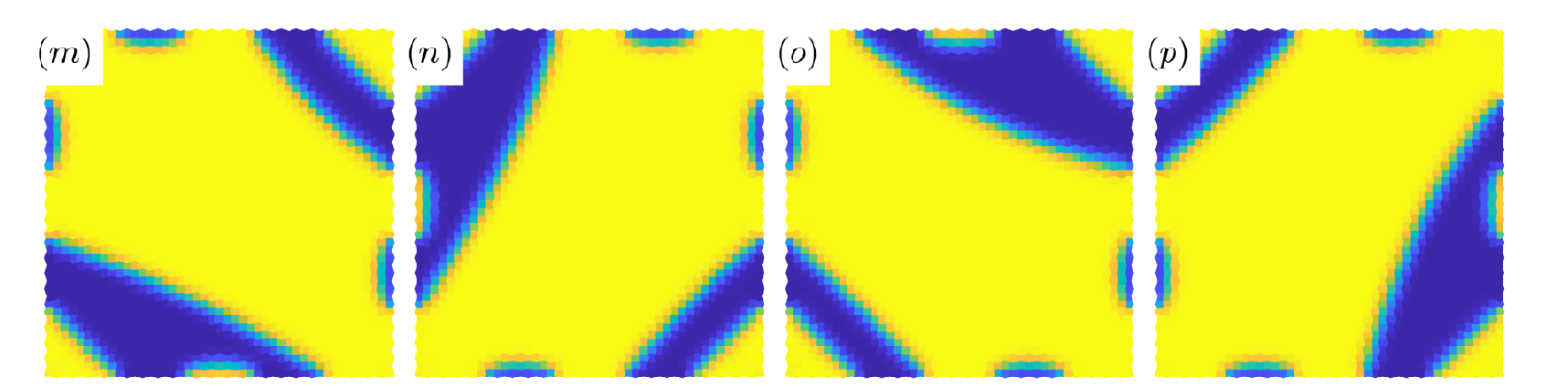}}
\end{subfigure}
\hfill
\caption{Degeneracy levels for configurations with $N=16$ alternating $M=\pm1$ segements evenly spaced along the boundary. All configurations in this square geometry will have degeneracy $1$, $2$, $4$, or $8$. Panels (a) and (b) are nondegenerate states. Panels (c) and (d) illustrate a state with degeneracy $2$. Panels (e)-(h) show $4$ degenerate states related by rotation. The remaining panels, (i)-(m) show an example of $8$ degenerate states.}
\end{figure}

It is possible to count the multiplicity of distinct metastable states as a function of the number $ N $ of alternating segments of orientation $A$ and $B$ along the boundary. Circling the boundary of the lattice, each defect marking the end of a domain wall is a transition from $A $ to $B$ or from $B$ to  $A$. Since each domain wall must both start and end on a boundary, each domain wall emanating from an $ AB$ defect must terminate at a $ BA $ defect, so that there are a total $N/2$ of each defect type. If there were an odd number of defects injected along the boundary, this would imply the existence of at least one domain wall in between that terminates in the bulk, which is impossible. Therefore, we know each $ AB $ transition is matched to a $BA$ transition, and we get a simple upper limit on the number of states by counting all permutations between the two sets: $ \Omega_{\rm max}(N) = (N/2)!$. 

This upper limit overestimates the number of states because it includes states in which different domain walls intersect, which is forbidden. This no-crossing restriction naturally leads to a recursion relation for the number of allowed configurations. Assume a boundary condition with $N+2$ total defects, and pick two of them to be connected. Since no additional domain wall can cross through this wall, the system is divided into two disjoint sections, one section containing $m=0,2,4,...N/2$ unassigned defects and the other section containing $N-m$ unassigned defects. Fixing the chosen defect pair, there are $\Omega(m)\Omega(N-m)$ possibilities. There are $N/2$ pairs of nodes that can be connected to define the two disjoint sections, and so that final number of topologically distinct states must satisfy the recursion relation \begin{equation} \Omega(N+2) = \sum_{m=0,2,4,...}^{N/2} \Omega(N-m)\Omega(m). \end{equation} Let the generating function associated with the number of states be defined as $g(z) = \sum_{n = 0}^{\infty} \Omega(2n)z^n$. For convenience, let $N=2y$ and $m=2x$, so that the recursion relation becomes  \begin{equation} \Omega[2(y+1)] = \sum_{x=0}^{y} \Omega[2(y-x)]\Omega(2x). \end{equation}  Multiplying both sides of the recursion relation by $z^{y+1}$, and then summing over $y$ gives \begin{equation} \sum_{y=0}^{\infty} \Omega[2(y+1)]z^{y+1} = \sum_{y=0}^{\infty} \sum_{x=0}^{y} \Omega[2(y-x)]\Omega(2x)z^{y+1}. \end{equation} Reordering the sums on the right side, the equation becomes \begin{equation} \sum_{y=0}^{\infty} \Omega[2(y+1)]z^{y+1} = z \sum_{x=0}^{\infty} \Omega(2x)z^{x} \sum_{y=x}^{\infty}  \Omega[2(y-x)]z^{y-x}. \end{equation} Invoking the generating function definition then gives \begin{equation} g(z)-1 = zg^2(z), \end{equation} and solving this quadratic relation for $g(z)$, we find that \begin{equation} g(z) = \frac{1 - \sqrt{1-4z}}{2z}, \end{equation} where the sign of the root is chosen so that $\lim_{z\rightarrow0} g(z) = \Omega(0) = 1$. Finally, the coefficients of the Taylor series expansion of $g(z)$ provide the exact multiplicity for any desired number of boundary segments $N$. The rapid growth of the number of possible configurations is shown in Fig.~11(g). We also checked that directly counting the topologically allowed $\Omega(N)$ up $N = 18$ matches the calculated formula. Illustrations of the resulting internal domain wall configurations are given in Fig.~11(a)-(f) as well as in Fig.~12. 

The generating function Eq.~E6 is in fact known to be the generating function for a well known series known as the Catalan numbers~\cite{wolfram}, $\Omega(2n) = C_{n} = \frac{1}{n+1} {2n\choose n}$. For $n$ large, the series grows asmptotically as $C_n \sim 4^n n^{-3/2}$.

{\color{black} Though the Catalan numbers account for the number of topologically distinct configurations, these configurations may be grouped by energy. The number of energetically degenerate states depends on the shape of the boundary, as well as on the spacing of the defects, both of which will determine the lengths of the internal domain walls. For example, Figure 11(a)-(f) shows that for $N = 8$ there are a total of $6$ energy levels. The remaining topologically distinct states are accounted for by $\pi/2$ rotations of the ones shown in Fig.~11. For a square boundary with $N/4$ defects on each side, states may be related by $\pi/2$ degree rotations, or by mirror reflection along the diagonal, so that all configurations for any $N$ have degeneracy $1, 2, 4$, or $8$. Examples of degenerate states for $N=16$ are given in Fig.~12. 

Classifying the configurations by energy becomes exceedingly difficult for increasing $N$. Figure 11 shows that for $N=8$, the $\Omega(8) = 14$ topological distinct states can be grouped into $6$ energetically degenerate states, $2\times1+2\times2+2\times4=14$. For $N = 12$, there are $\Omega(12) = 132$, and we have checked these may be grouped into $48$ energetically distinct states, $8 \times 1 + 18\times2 + 22\times4 = 132$. For the case $N = 16$, $\Omega(16) = 1,430$, and enumerating the energy levels becomes exceedingly difficult. Since each state has a maximum degeneracy $8$, we obtain a lower bound of $1,430/8\approx178$ distinct configurations. As a check, we have simulated an ensemble of $2,000$ trials, using $N=16$, $L_x=L_y=45$, and $k_2=0.5$. In each case we start from a random state, and minimize the energy by gradient descent. Sorting the results of these runs by energy, we find $63$ energy levels, see Fig.~13 and Fig.~14, where the states are presented in order of increasing energy. Counting the possible rotations and reflections of these $63$ groups, gives a total of $381$ out of the expected $1,430$ configurations. The undiscovered states presumably have higher energies, making them difficult to generate via our energy minimization protocol.}

\begin{figure}[b!]
\begin{subfigure}{\linewidth}
\resizebox{\linewidth}{!}{\includegraphics{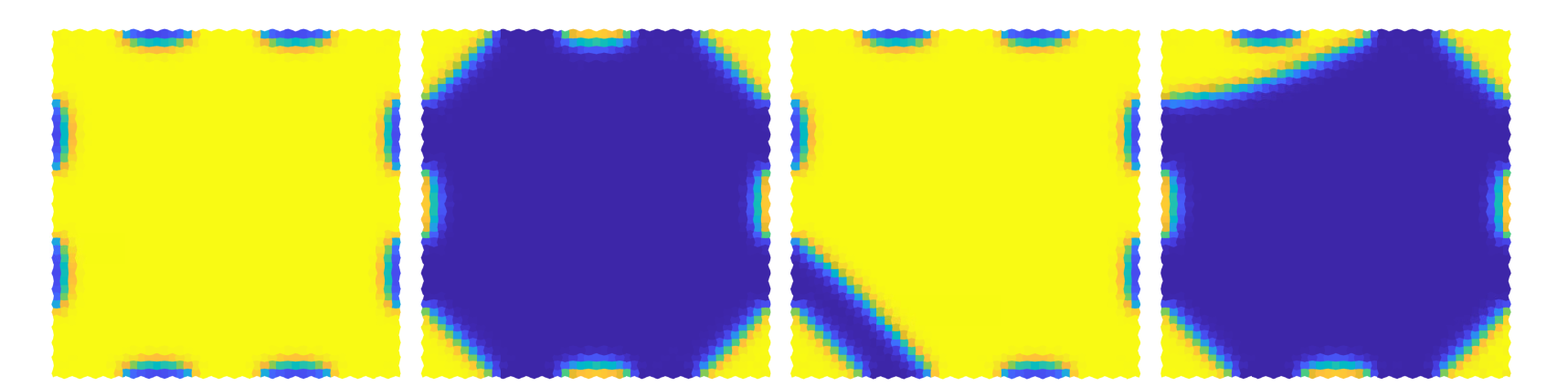}}
\end{subfigure}
\begin{subfigure}{\linewidth}
\resizebox{\linewidth}{!}{\includegraphics{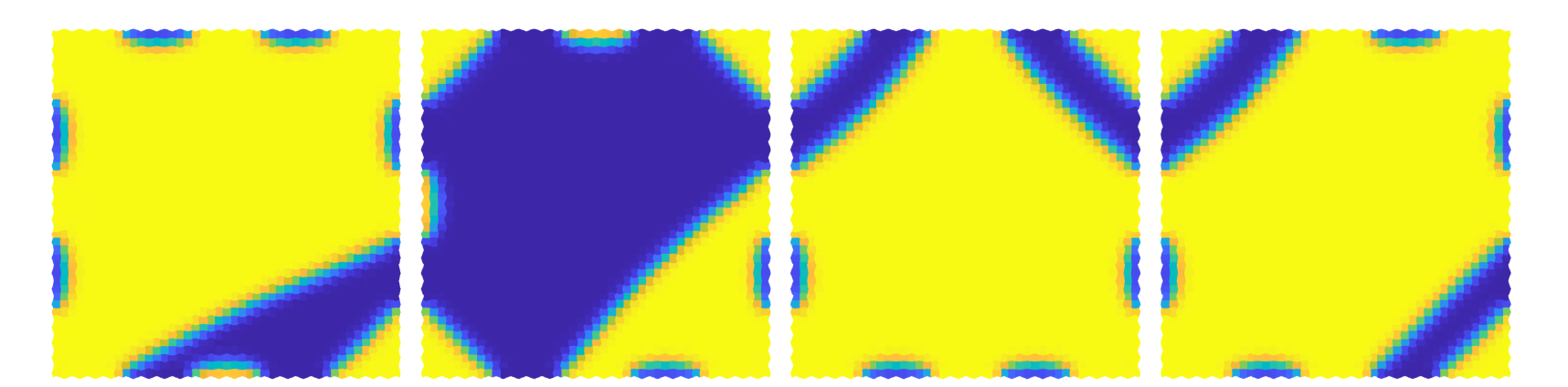}}
\end{subfigure}
\begin{subfigure}{\linewidth}
\resizebox{\linewidth}{!}{\includegraphics{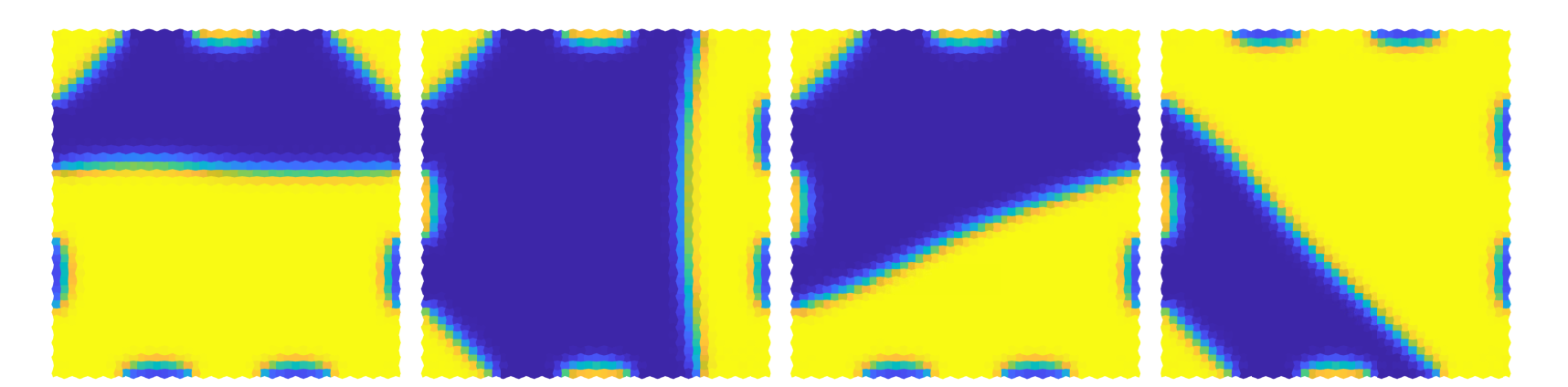}}
\end{subfigure}
\begin{subfigure}{\linewidth}
\resizebox{\linewidth}{!}{\includegraphics{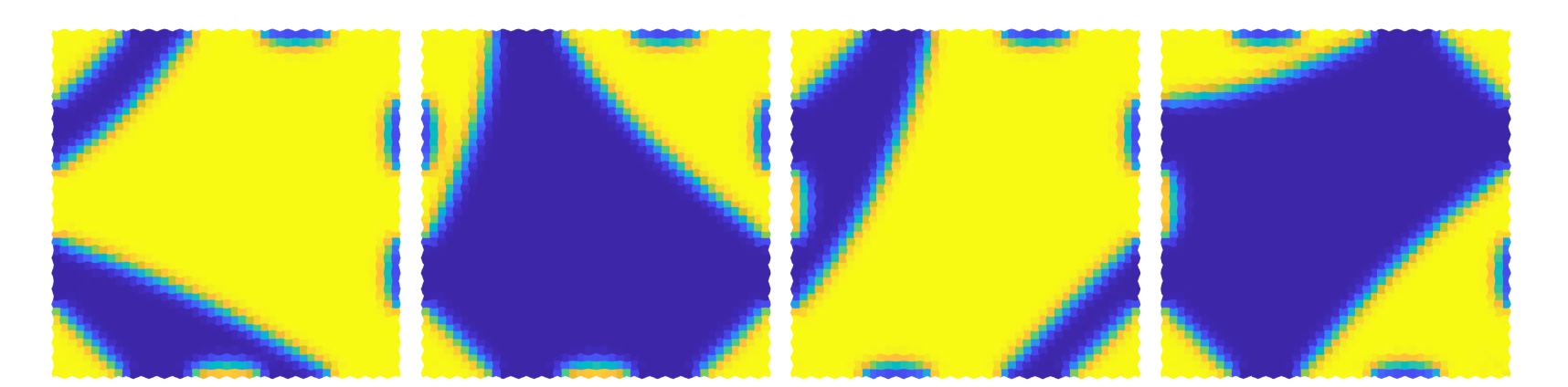}}
\end{subfigure}
\begin{subfigure}{\linewidth}
\resizebox{\linewidth}{!}{\includegraphics{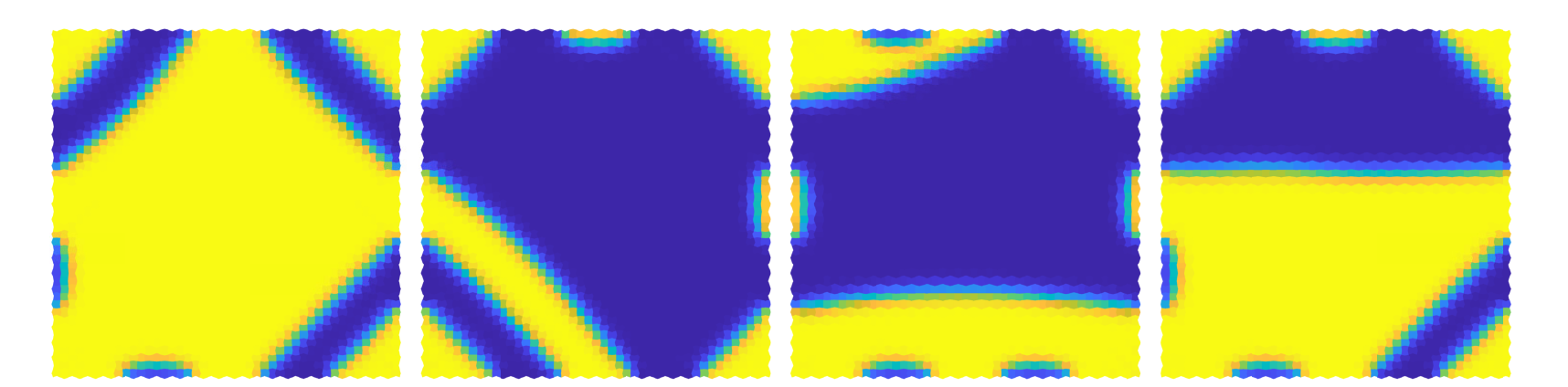}}
\end{subfigure}
\begin{subfigure}{\linewidth}
\resizebox{\linewidth}{!}{\includegraphics{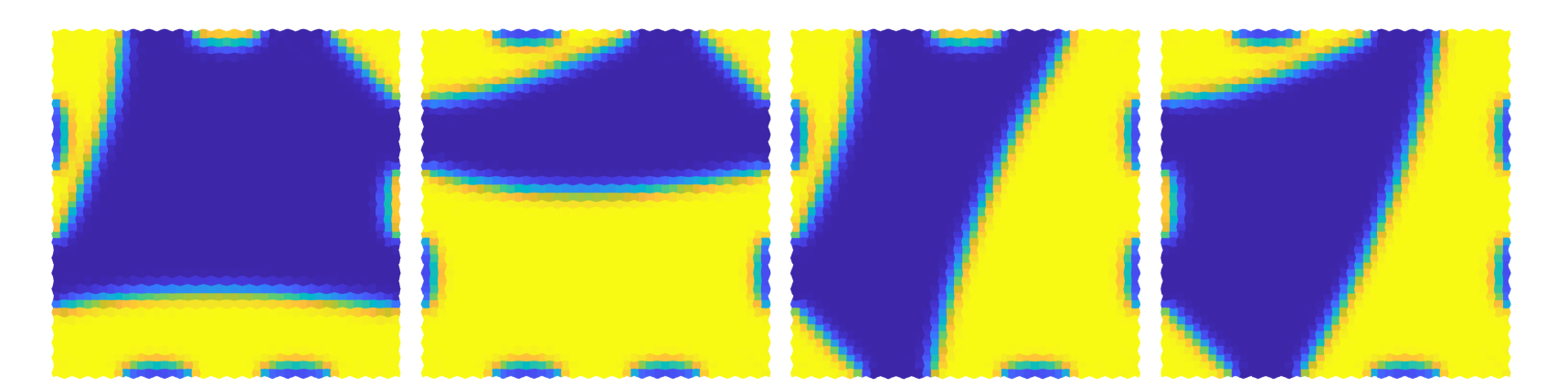}}
\end{subfigure}
\begin{subfigure}{\linewidth}
\resizebox{\linewidth}{!}{\includegraphics{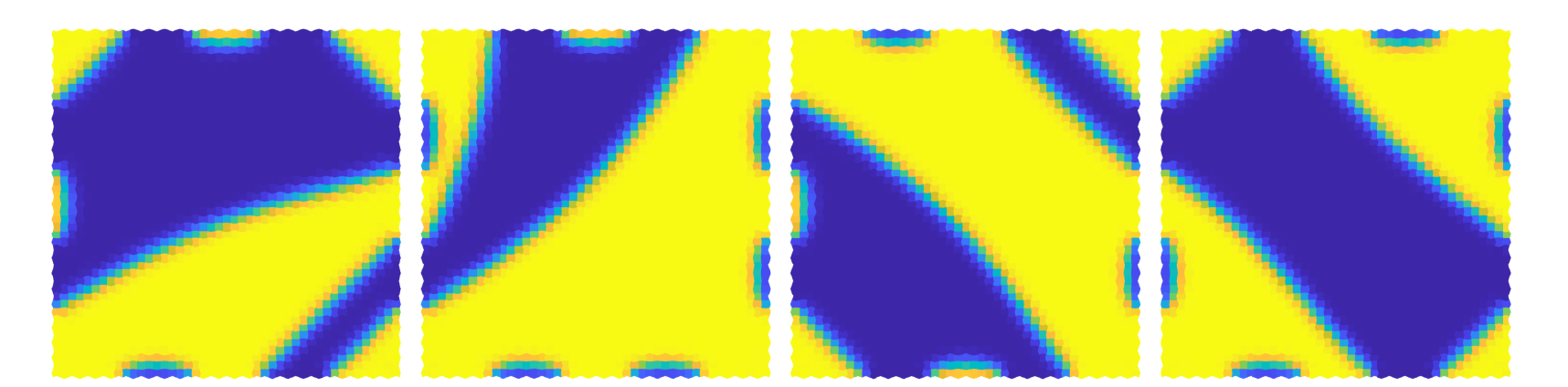}}
\end{subfigure}
\begin{subfigure}{\linewidth}
\resizebox{\linewidth}{!}{\includegraphics{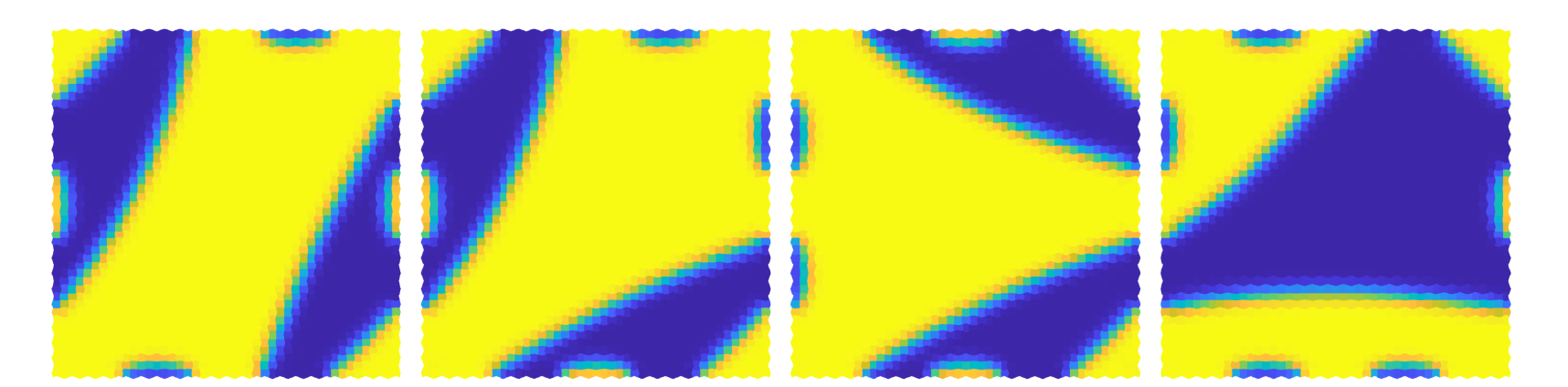}}
\end{subfigure}
\hfill
\caption{First $32$ numerically sampled distinct metastable states for $N=16$ evenly spaced boundary defects, system size $L_x = L_y = 45$ and interaction spring stiffness $k_2 = 0.5$. The configurations are displayed in order of increasing total energy, from top left to bottom right.}
\end{figure}

\begin{figure}[b!]
\begin{subfigure}{\linewidth}
\resizebox{\linewidth}{!}{\includegraphics{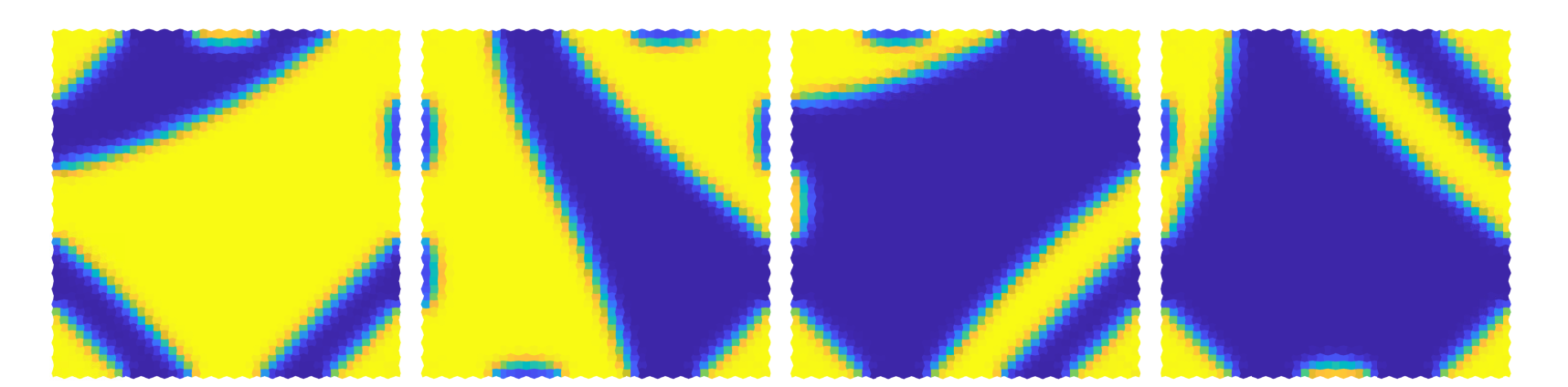}}
\end{subfigure}
\begin{subfigure}{\linewidth}
\resizebox{\linewidth}{!}{\includegraphics{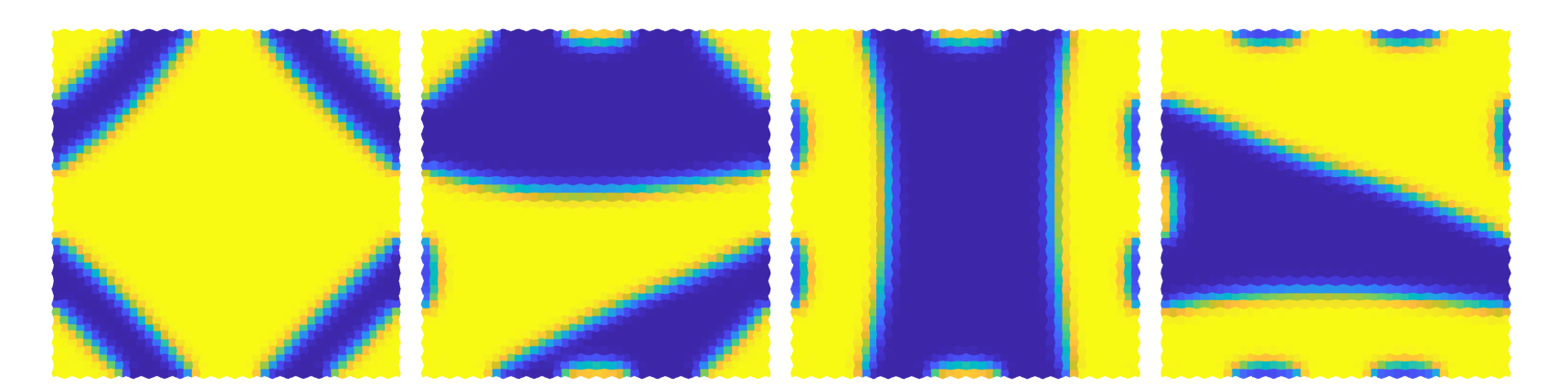}}
\end{subfigure}
\begin{subfigure}{\linewidth}
\resizebox{\linewidth}{!}{\includegraphics{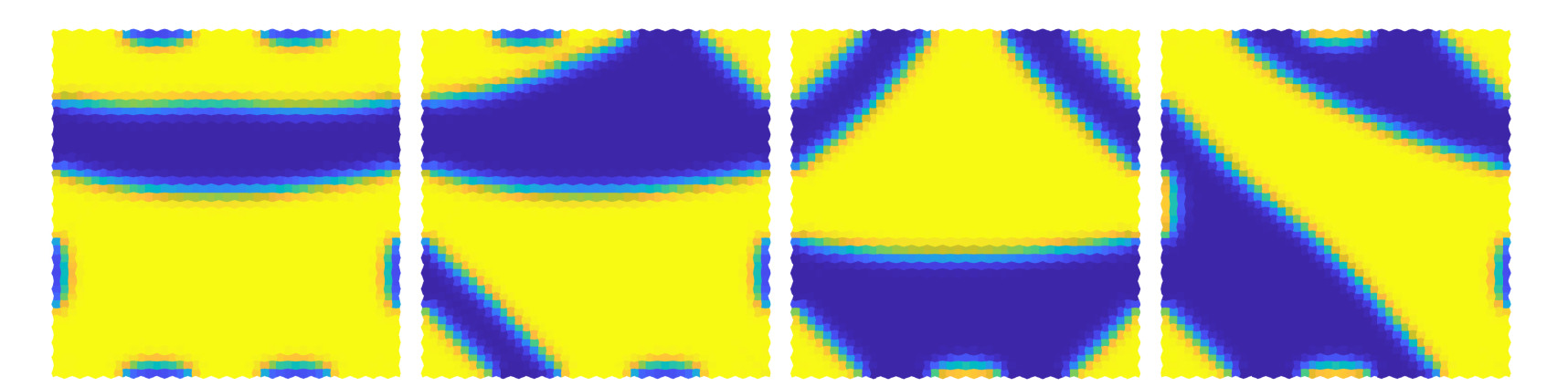}}
\end{subfigure}
\begin{subfigure}{\linewidth}
\resizebox{\linewidth}{!}{\includegraphics{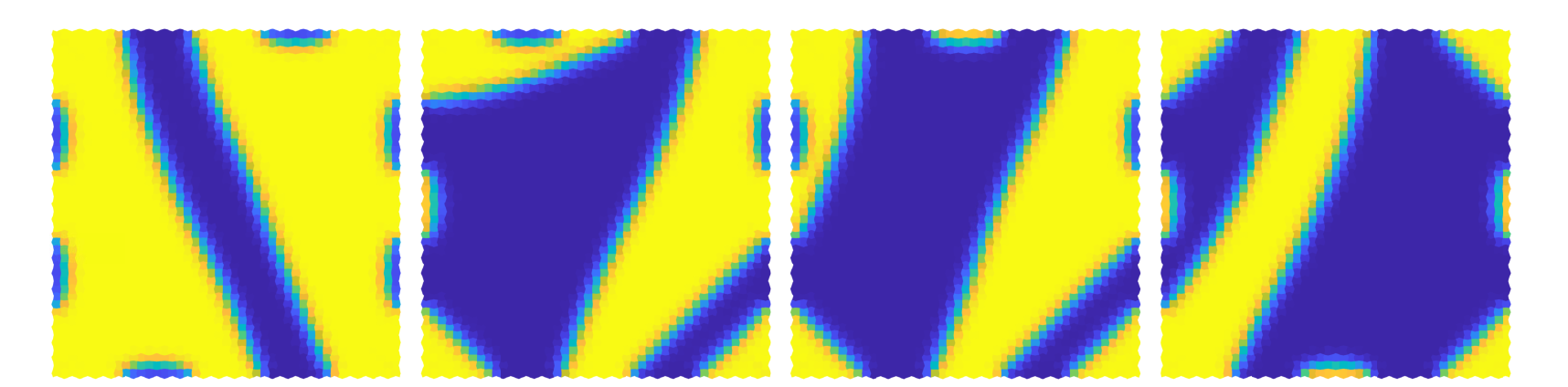}}
\end{subfigure}
\begin{subfigure}{\linewidth}
\resizebox{\linewidth}{!}{\includegraphics{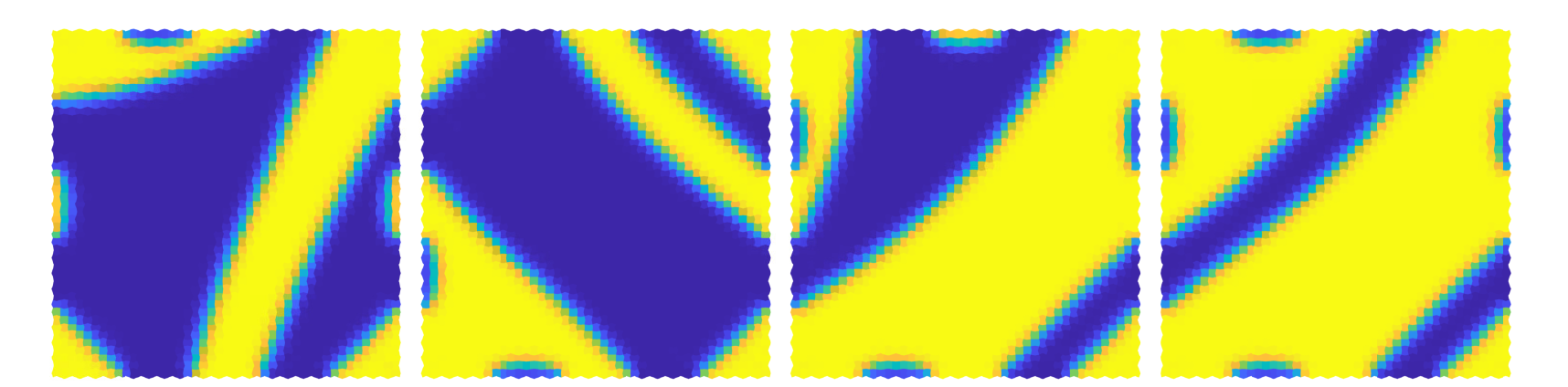}}
\end{subfigure}
\begin{subfigure}{\linewidth}
\resizebox{\linewidth}{!}{\includegraphics{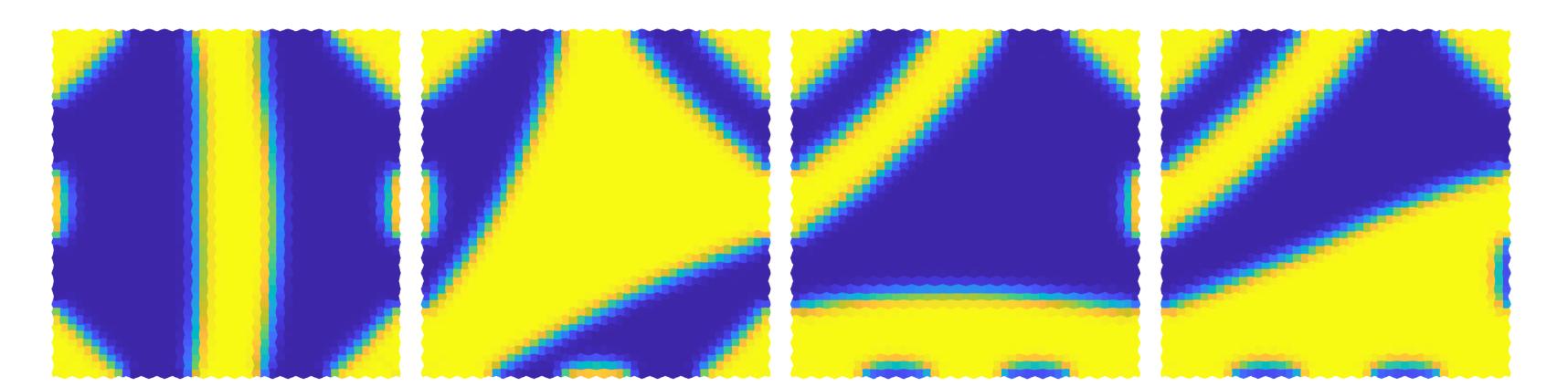}}
\end{subfigure}
\begin{subfigure}{\linewidth}
\resizebox{\linewidth}{!}{\includegraphics{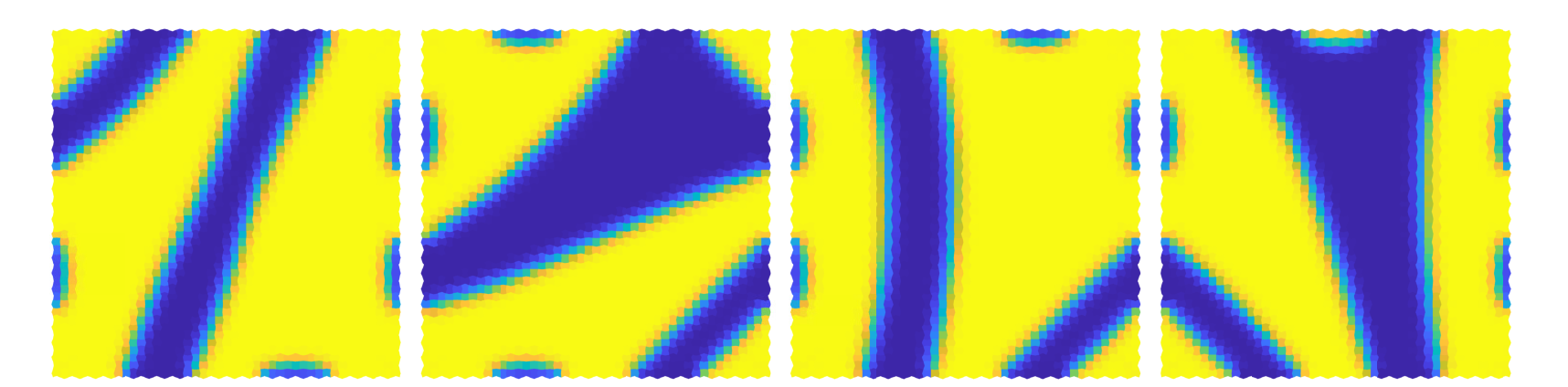}}
\end{subfigure}
\begin{subfigure}{\linewidth}
\resizebox{\linewidth}{!}{\includegraphics{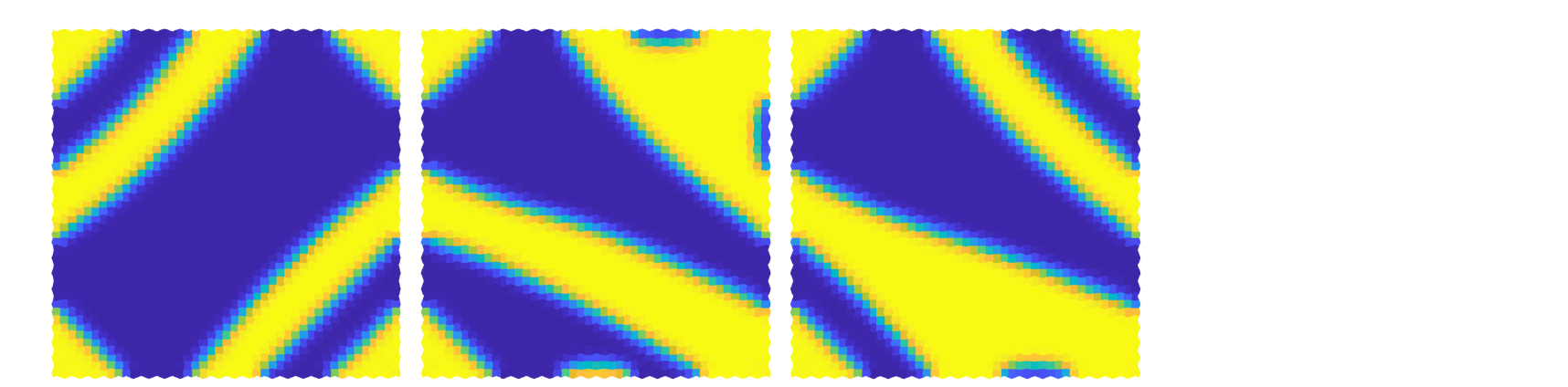}}
\end{subfigure}
\hfill
\caption{Additional $31$ numerically sampled metastable states for $N=16$ evenly spaced boundary defects, system size $L_x = L_y = 45$ and interaction spring stiffness $k_2 = 0.5$. The configurations are displayed in order of increasing total energy, from top left to bottom right.}
\end{figure}

\clearpage

\bibliography{Metamaterials_References,Spin_Ice_References,Dynamic_Hysteresis_References}

\end{document}